\documentclass[10pt]{amsart}

\title{Decidability and Universality in Symbolic Dynamical Systems}

\author{Jean-Charles Delvenne}
\address{Universit\'{e} catholique de Louvain,
Department of Mathematical Engineering \\
Avenue Georges Lema\^{i}tre 4, B-1348 Louvain-la-Neuve, Belgium}
\email{delvenne@inma.ucl.ac.be}
\author{Petr K\r{u}rka}
\address{Center for Theoretical Study $\;\&\;$
Faculty of Mathematics and Physics\\
Charles University in Prague, Malostransk\'{e}
n\'{a}m\v{e}st\'{\i} 25, CZ-11800 Praha 1, Czechia}
\email{kurka@ms.mff.cuni.cz}
\author{Vincent D. Blondel}
\address{Universit\'{e} catholique de Louvain,
Department of Mathematical Engineering \\
Avenue Georges Lema\^{i}tre 4, B-1348 Louvain-la-Neuve, Belgium}
\email{blondel@inma.ucl.ac.be}

\newcommand{\R}{\mathbb R}
\newcommand{\N}{\mathbb N}
\newcommand{\Z}{\mathbb Z}
\newcommand{\Aa}{\mathcal A}
\newcommand{\Bb}{\mathcal B}
\newcommand{\Ff}{\mathcal F}
\newcommand{\Ll}{\mathcal L}
\date{}

\usepackage{indentfirst}

\usepackage{array}

\usepackage{latexsym}
\usepackage{amssymb}
\usepackage{amsmath}
\usepackage[dvips]{graphicx}

\usepackage{pictex}

\usepackage[latin1]{inputenc}

\newtheorem{defi}{Definition}
\newtheorem{exam}{Example}
\newtheorem{prop}{Proposition}
\newtheorem{lemm}{Lemma}
\newtheorem{conj}{Conjecture}

\newtheorem{coro}{Corollary}

\begin{document}

\maketitle

\begin{abstract}
Many different definitions of computational universality for
various types of dynamical systems have flourished since Turing's
work. We propose a general definition of universality that applies
to arbitrary discrete time symbolic dynamical systems.
Universality of a system is defined as undecidability of a
model-checking problem. For Turing machines, counter machines and
tag systems, our definition coincides with the classical one. It
yields, however, a new definition for cellular automata and
subshifts. Our definition is robust with respect to initial
condition, which is a desirable feature for physical
realizability.

We derive necessary conditions for undecidability and
universality. For instance, a universal system must have a
sensitive point and a proper subsystem. We conjecture that
universal systems have infinite number of subsystems. We also
discuss the thesis according to which computation should occur at
the `edge of chaos' and we exhibit a universal chaotic system.
\end{abstract}


\section{Introduction}

Computability is usually defined via universal Turing machines. A
Turing machine can be regarded as a dynamical system, i.e., a set of
configurations  together with a transformation acting on this set.
A configuration consists of the state of the head and the
content of the tape. Computation is done by observing the
trajectory of an initial point under iterated transformation.

There is no reason why Turing machines should be the only
dynamical systems capable of universal computation.  Indeed, such
capabilities have been also claimed for artificial neural networks
\cite{siegelmann,koiran_recur_networks_96}, piecewise linear maps
\cite{koiran_piecewise_lin_maps_94}, analytic maps
\cite{koiran_moore_99}, cellular automata \cite{wolfram},
piecewise constant vector fields \cite{asarin_pnu_maler_95},
billiard balls on particular pool tables
\cite{Fredk_Toffo_billiard}, or a ray of light between a set of
mirrors \cite{moore_unpred_and_undec}. For all these systems, many
particular definitions of universality have been proposed. Most of
them mimic the definition of computation for Turing machines: an
initial point is chosen, then we observe the trajectory of this
point and see whether it reaches some `halting' set; see for
instance \cite{siegelmann__fishman_analog_dissipative} and
\cite{bournez_cosnard}. However, many variants of these
definitions are possible and lead to different classes of
universal dynamical systems. In particular, there is no consensus
for what it means for a cellular automaton to be universal.
Moreover, in the presence of noise many of these systems loose
their computational properties
\cite{asarin_bouajjani,maass_orponen,gacs97reliable}; see
\cite{orponen_survey,moore_auckland,moore_recogn} for definitions
of analog computation and issues relative to noise and physical
realizability.

Another field of investigation is to make a link between the
computational properties of a system and its dynamical properties.
For instance, attempts have been made to relate `universal'
cellular automata to Wolfram's classification. It has also been
suggested that a `complex' system must be on the `edge of chaos':
this means that the dynamical behavior of such a system is neither
simple (i.e., a globally attracting fixed point) nor chaotic; see
\cite{wolfram,edgeofchaos_reexamination,crutchfield_onsetchaos,langton_edgeofchaos}.
Other authors nevertheless argue that a universal system may be
chaotic; see \cite{siegelmann}.

The basic questions we would like to address are the following:
\begin{itemize}
    \item How to define computationally universality for dynamical systems?
    \item What are the dynamical properties of  a universal system?
\end{itemize}

A long-term motivation is to answer these questions from the point
of view of physics. What physical systems are universal? Is the
gravitational N-body problem universal
\cite{moore_unpred_and_undec}? Is the Navier-Stokes equation
universal \cite{moore_generalizedshifts}? However in this paper we
focus on \emph{symbolic effective} dynamical systems, i.e.,
systems  defined on the Cantor set $\{0,1\}^\N$ or a subset of it,
whose transformations are computable. Some motivating examples of
such systems are Turing machines, cellular automata and subshifts.

Turing's machine was originally meant as a model of a computation
performed by a human operator using paper and pencil
\cite{turing}. We adapt Turing's reasoning to the case where the
human operator does not compute by himself, but relies on a
dynamical system to make the computation. The system is said to be
computationally universal if the observations made by the human
operator allow him to solve any problem that could also be solved
by a universal Turing machine. We conclude that a system is
universal if some property of its trajectories, such as
reachability of a halting set, is r.e.-complete. This is an
extension of Davis' definition of universal Turing machine.

In this contribution, rather than considering point-to-point or
point-to-set properties, we consider set-to-set properties.
Typically, given an initial set and a halting set, we want to know
whether there is at least one configuration in the initial set
whose trajectory eventually reaches the halting set. We require
the initial and halting sets to be clopen (closed and open) sets
of the Cantor state space. Clopen sets can be described in a
natural way with a finite number of bits. Finally, we do not
restrict ourselves to the property `Is there a trajectory going
from $U$ to $V$?' alone. In a previous paper
\cite{delvenne_kurka_blondel} we have considered properties
expressible by temporal logic. In the present paper we consider
the wider class of all properties that can be observed by some
finite automaton. Checking whether such a property is verified is
a `model-checking problem'. Model-checking problems are usually
defined for finite or countable systems; see
\cite{modelchecking1999} for instance. Finally, a universal system
is a system that has an r.e.-complete model-checking problem.

This definition addresses the two issues raised above. Firstly, it
is a general definition directly applicable to any (effective) symbolic
system. Secondly, dealing with clopen sets rather than points takes
into account some constraints of physical realizability, such as
robustness to noise.
With this definition in mind, we prove necessary conditions for a
symbolic system to be universal. In particular, we show that a
universal symbolic system is not minimal, not equicontinuous and
does not satisfy the shadowing property. We conjecture that a
universal system must have infinitely many subsystems, and we show
that there is a chaotic system that is universal, contradicting
the idea that computation can only happen at the `edge of chaos'.

Preliminaries are given in Sections \ref{basics}, \ref{basics2}
and \ref{sectautomata}. Decidable and universal systems  are
defined in Sections \ref{sectdecidability} and
\ref{sectuniversality}. In Section
\ref{sectnecesscond}, necessary conditions for a system to be
universal are given, related to minimality, equicontinuity and
shadowing property; chaos and edge of chaos are considered in
Section \ref{sectunivchaos}. The definition of universality is
discussed in Section \ref{sectdiscussdef}.

\section{Effective symbolic spaces} \label{basics}

A \emph{symbolic space} is a compact metric space whose clopen
(closed and open) sets form a countable basis: every open set is a
union of clopen sets. The elements of a symbolic space are called
\emph{points} or \emph{configurations}. A typical example is the
Cantor set $\{0,1\}^\N$ endowed with the product topology. The
topology is given by the metric $d(x, y)= 2^{-n}$, where $n$ is
the index of the first bit on which $x$ and $y$ differ. Note that
this metric satisfies the \emph{ultrametric inequality}: $d(x,z)
\leq \max(d(x,y),d(y,z))$ for any $x$, $y$ and $z$.

If $w \in \{0,1\}^*$ is a finite binary word, then $[w]$ denotes
the set of all infinite  configurations with prefix $w$. Sets of
this form, usually called \emph{cylinders}, are exactly the balls
of the metric space. They are clopen sets and any clopen set of
$\{0,1\}^\N$ is a finite union of cylinders. Similar distances are
defined on the spaces $\{0,1\}^* \cup \{0,1\}^\N$, $A^\N$, $Q
\times A^\Z$, $A^{\Z^d}$ where $Q$ and $A$ are finite and $d$ is a
positive integer. Closed subsets of the Cantor space are symbolic
spaces themselves. The converse is well known to hold as well:
Every symbolic space is homeomorphic to a closed subset of the
Cantor space and every perfect symbolic space is homeomorphic to
the Cantor space. For instance, $\{0,1\}^\Z$ is homeomorphic to
$\{0,1\}^\N$.

To define computational universality, we need effective
symbolic spaces, in which we can perform boolean combinations
on clopen sets effectively.

\begin{defi} \label{defeffspace}
An \emph{effective symbolic space} is a pair $(X, P)$, where $X$
is a symbolic space  and  $P:\N \rightarrow 2^X$ is an injective
function whose range is the set of all clopen sets of $X$, such
that the intersection and complementation of clopen sets are
computable operations. This means that there exist computable
functions $f:\N \to \N$ and $g:\N\times \N \to \N$ such that
$X\setminus P_n = P_{f(n)}$ and $P_n \cap P_m = P_{g(n,m)}$.
\end{defi}
Of course, union of clopen sets is then also computable. Often we
denote an effective symbolic space by $X$ rather than $(X, P)$
when no confusion is to be feared. In Cantor space $\{0,1\}^\N$,
the lexicographic ordering yields a standard enumeration
$$ [\lambda],[0],[1],[00],[01],[10],[11], [00]\cup[11], [01]\cup[10],
 [00]\cup[01]\cup[10], [00]\cup[01]\cup[11],\ldots $$
Other widely used symbolic spaces like $\{0,1\}^* \cup \{0,1\}^\N$,
$A^{\N^d}$, $A^{\Z^d}$, $Q \times A^\Z$, have also their standard
enumerations. Note that we could require intersections and complements to be
primitive recursive rather than computable, without altering
the examples and results of the paper.

\begin{defi} \label{defeffcont}
Let $(X,P)$ and $(Y,Q)$ be two effective symbolic spaces. An
\emph{effective continuous map} is a continuous map $h:X
\rightarrow Y$ such that $h^{-1}(Q_n)= P_{k(n)}$, for some
computable map $k: \N \rightarrow \N$.
If $h$ is bijective then it is an \emph{effective homeomorphism},
and $(X,P)$ is said to be \emph{effectively homeomorphic} to
$(Y,Q)$.
\end{defi}

Note  that the composition of effective continuous maps is an
effective continuous map, the identity is an effective continuous
map and the inverse map of an effective homeomorphism is also an
effective homeomorphism.  In particular, being effectively
homeomorphic is an equivalence relation for effective symbolic
spaces.

Given an effective symbolic space $(X, P)$, a closed subset $Y$ is
said to be \emph{effective}, if the family of clopen sets
intersecting $Y$ is decidable. In particular any clopen set is
effective. An effective set $Y$ can be endowed with the relative
topology, whose clopen sets are all intersections of clopen sets
of $X$ with $Y$. Thus, the enumeration $P_0,P_1,P_2,\ldots$ of
clopen sets of $X$ yields an enumeration of clopen sets of $Y$:
$Y \cap P_0$, $Y \cap P_1$, $Y \cap P_2$, \ldots. This enumeration
may contain empty sets and repetitions, but we can detect them
in an effective way and renumber the sequence accordingly.
Hence we get an effective topology for the effective closed
set $Y$. Equivalently, the
inclusion $i: Y \hookrightarrow X$ is an effective continuous map.

\begin{prop} \label{propall is eff cantor}
Every effective symbolic space is effectively homeomorphic to an
effective subset of the Cantor space. Every perfect effective
symbolic space is effectively homeomorphic to the Cantor space.
\end{prop}

\begin{proof} Let $(X,P)$ be an effective symbolic space.
For every point $x \in X$, construct the infinite configuration
$g(x) \in \{0,1\}^\N$, where $g(x)_n=1$ if and only if $x \in
P_n$. Then the map $g: X \rightarrow \{0,1\}^\N$ is injective and
continuous. Since $X$ is compact, $g(X)$ is closed. Moreover,
every step of the construction is effective, and so $g(X)$ is an
effective closed set and the map $g$ is effective.

If the space is perfect, then we construct another map $h:X \to
\{0,1\}^\N$. We may write $X$ as a partition of two clopen sets
$X=A_0 \cup A_1$, where $A_0$ is the clopen set of smallest index
to be different from $X$ and $\emptyset$; this is always possible
thanks to perfectness. Suppose that we have already constructed
$A_w$, where $w$ is a binary word. Let $n$ be the first index such
that $A_w \cap P_n$ differs from both $A_w$ and $\emptyset$, and
set $A_{w0} = A_w \cap P_n$, $A_{w1} = A_w\setminus P_n$. For $x
\in X$ let $h(x) \in \{0,1\}^{\N}$ be the unique configuration
such that $x\in A_w$ for all prefixes $w$ of $h(x)$. Then $h:X \to
\{0,1\}^{\N}$ is an effective homeomorphism.
\end{proof}

We see that there is no loss of generality in supposing that
in any effective symbolic space, for any rational $\epsilon$ there
exists a finite number of balls of radius  $\epsilon$ and that we can
compute all of them. Indeed, this is the case for all effective
subsets of the Cantor space.

\section{Effective symbolic systems}  \label{basics2}


\begin{defi}
An \emph{effective symbolic dynamical system} is an effective
continuous map from an effective symbolic space to itself.
\end{defi}

In other words, an effective symbolic system is a symbolic space
with a continuous self-map in which intersections, complements,
and inverse images of clopen sets are computable. This definition
of effective function in a Cantor space is equivalent to classical
definitions in computable analysis; see for instance
\cite{weihrauch}.
We denote an effective symbolic system by a map $f:X \rightarrow X$
or simply $f$, when the enumeration $P$ of $X$ is implicit.
Extending Definition \ref{defeffcont}, we define a relation of
equivalence between effective systems.

\begin{defi}
The effective symbolic systems $f:X \rightarrow X$ and $g:Y
\rightarrow Y$ are \emph{effectively conjugated} if there exists an
effective homeomorphism $h: X \rightarrow Y$ such that $h \circ f
= g \circ h$. If $h: X \rightarrow Y$ is an effective surjective map
(and not bijective), then the system $g:Y \rightarrow Y$ is said
to be an \emph{effective factor} of $f:X \rightarrow X$.
The factor $g$ can be seen as a `simplification' of $f$.
\end{defi}

The identity on any symbolic space is the simplest example of an
effective symbolic system.
A cellular automaton is an effective symbolic system acting on the
space $A^{\Z^d}$, where $A$ is the finite alphabet and $d$ is the
dimension.

Turing machines are usually described as working on finite
configurations. A finite configuration is an element of
$\{0,1\}^* \times Q \times \{0,1\}^*$, where $Q$ denotes
the set of states
of the head, the first binary word is the content of the tape to the
left of the head and the second binary word is the right part of
the tape. However, $\{0,1\}^*$ cannot be naturally equipped with
a compact topology, so we consider its compactification
 $W=\{0,1\}^* \cup \{0,1\}^\N$, i.e., the set of finite and infinite
binary words. Then the Turing machine function is also defined on
$W \times Q \times W$, which is a compact space, whose isolated
points are $\{0,1\}^*\times Q \times \{0,1\}^*$. An isolated point
is clopen in $W\times Q \times W$.
Hence a Turing machine with a blank symbol is an effective symbolic
system on the space $W \times Q \times W$.

A Turing machine without blank symbol is an effective symbolic
system as well. As we do not suppose that almost all cells are
filled with a blank symbol, a configuration is given by an
arbitrary element of $\{0,1\}^\N \times Q \times \{0,1\}^\N$
or, equivalently, $Q \times A^{\Z}$. This is a Turing machine with
moving tape, as considered in \cite{kurka_topo_TM}: the head is
always in position zero, and the tape moves to the left or to the
right.

\subsection{Shifts and subshifts}
\label{subsect shifts and subshifts}

A one-sided or two-sided \emph{shift} is a dynamical system on
$A^\N$ or $A^\Z$ (where $A$ is a finite alphabet) with the map
$\sigma: A^\N \rightarrow A^\N$ or
$\sigma: A^\Z \rightarrow A^\Z$ defined by $\sigma(x)_i = x_{i+1}$.
A shift is an effective system.
A \emph{subshift} is a subsystem of the shift, i.e., a closed
subset that is invariant under the shift map. Most subshifts we
consider in this article are one-sided subshifts.
An \emph{effective subsystem} of an effective symbolic system is
an effective closed subset that is invariant under the map. With
the relative topology, it is itself an effective symbolic system.
In particular, a subshift that is an effective closed subset of
$A^\N $ is again an effective symbolic system.

The set $\Ll(X)$ of all finite words appearing at least once in at
least one point of the subshift $X$ is called the \emph{language}
of the subshift. It is easy to see that a subshift is effective
iff its language is recursive. In particular every \emph{sofic}
subshift (a subshift whose language is regular) is effective. A
subshift of finite type is the set of sequences avoiding a finite
set of forbidden subwords. Subshifts of finite type are sofic
subshfits, hence are effective. Another widely studied class of
subshifts are \emph{substitutive} subshifts defined by
substitutions $\vartheta: A \to A^+$. Since a substitution is a
finitary object, every substitutive subshift is effective. A
Sturmian subshift $\Sigma_{\alpha}$ associated to an irrational
number $\alpha$ is a symbolic model of rotation of the circle $x
\mapsto x+\alpha$; see e.g. \cite{kurka_book}. A Sturmian subshift
$\Sigma_{\alpha}$ is is effective iff $\alpha$ is a computable
real number.

From any symbolic dynamical system (effective or not), we can
generate one-sided subshifts in a natural way. A \emph{clopen
partition} of a symbolic space is a partition of the space into a
finite number of disjoint clopen sets. A partition $\Aa$ is
\emph{finer} than $\Bb$, or $\Bb$ is \emph{coarser} than $\Aa$,
if every clopen set of $\Aa$ is included in some clopen set of
$\Bb$. Given a clopen partition $\Aa=\{A_1,\ldots,A_N\}$ of $X$,
the subshift \emph{induced} by this partition is the set of
infinite words $a_0a_1a_2a_3\ldots \in \Aa^\N$, such that there is
a point in $a_0$ whose trajectory goes successively through $a_1$,
$a_2$, \ldots. Note that here $A_1$, say, is both a subset of $X$
and a symbol from a finite alphabet. Thus $A_1 A_3 A_1$ denotes a
word of three symbols and not for instance a cartesian product.
The language of the subshift is also said to be \emph{induced} by
the partition. An induced subshift is a factor of the system and
conversely any factor subshift is induced by a clopen partition.
Following  this observation, we can characterize effective
symbolic systems in terms of their induced subshifts.

\begin{prop}
A symbolic system is effective if and only if there is an
algorithm deciding from any given clopen partition and any given
finite word whether this word belongs to the language of the
subshift induced by the partition.
\end{prop}

\begin{proof} Let $\Aa=\{A_1,\ldots,A_N\}$ be a clopen partition. Then a
word $a_0a_1\ldots a_{l-1} \in \Aa^*$ is in the language of the
subshift induced by the partition if and only if $a_0 \cap
f^{-1}(a_1) \cap \cdots \cap f^{-(l-1)}(a_{l-1})$ is not empty.
But this can be checked algorithmically.

Conversely, suppose that all induced subshifts have decidable
languages, and that given the partition  we can effectively find a
decision algorithm for the corresponding language. Let $P_n$ be a
clopen set of $X$. There exists a clopen partition
$\Aa=\{A_1,\ldots,A_N\}$ such that
\begin{itemize}
 \item for every $i$, either $A_i \subseteq P_n$ or
    $A_i \subseteq X \setminus P_n$;
 \item if $A_i A_j$ and $A_i A_k$ belong to the language of the
    induced subshift, then $A_j$ and $A_k$ are either both parts of
    $P_n$ or both parts of $X \setminus P_n$.
\end{itemize}
The first condition says that the partition is finer then $P_n$,
the second condition says that the partition is finer than
$f^{-1}(P_n)$. It can be checked algorithmically whether a clopen
partition has these two properties. Thus a partition with these
properties can be found algorithmically. Then we can compute
$f^{-1}(P_n)$ as the union of all $A_i$ such that there exists a
word $A_i A_j$ in the language of the induced subshift and that
$A_j \subseteq P_n$.
\end{proof}

If the subshifts have decidable languages, but decision algorithms
are not computable with respect to the clopen partition, then the
system may fail to be effective. This happens in the following
example.

\begin{exam}
Assume $k:\N \rightarrow \N$ is  an non-computable  strictly
increasing total function. We define a function $f$ on the Cantor
space $\{0,1\}^\N$ by $f(x)=f_0(x)f_1(x)f_2(x)\ldots$, where
$f_i(x_0 x_1 x_2 \ldots) = \max \{x_0, x_1, x_2, \ldots,
x_{k(i)}\}$. There are two fixed points, $0^\omega$ and
$1^\omega$, and the image of a point is of the form $0^*1^\omega$
or $0^\omega$ (where $0^\omega$ is a shortcut for $000\ldots$).
Then it is easy to see that for any point $x$ either
$f(x)=0^\omega$ or $f^n(x) = 1^\omega$ for some $n\geq 0$ . For
any partition $\Aa=\{A_1,\ldots,A_N\}$, if $0^\omega \in A_1$ and
$1^\omega \in A_2$ (say), then every point in $A_3 \cup \ldots
\cup A_N$ reaches $A_2$ in bounded time, say $t$. Then every
finite word of the language of the subshift induced by the
partition is of the form $A_1^*$ or $SA_2^*$, where $S$ is some
subset of $\{A_1,\ldots,A_N\}^t$. This is certainly a decidable
language. However $f$ is not effective, for otherwise we could
compute $k$.
\end{exam}

In the rest of the paper, we use the terms `symbolic system' or
even `system' to denote an effective symbolic dynamical system.

\subsection{Products}


Let $(f_n: X_n \rightarrow X_n)_{n \in \N}$ be a family  of
\emph{uniformly effective} systems on the effective symbolic
spaces $(X_n, P_n)$ ; we mean that there exists an algorithm that,
given $n$ and two clopen sets of $X_n$, can compute their
intersection, complements and inverse images.
Then the \emph{effective product} of $(f_n)_{n \in \N}$ is  the
system $f: X \rightarrow X$ on the effective symbolic space
$(X,P)$ such that
\begin{itemize}
    \item the set $X$ is the product of all sets $X_n$;
    \item the clopen sets of $X$ are all products of clopen sets
  $\prod_{n\in \N} A_n$ such that only finitely many $A_n\subseteq X_n$
   are different from $X_n$ (this is the usual product topology);
    \item the clopen sets are
        indexed by finite sets of integers in a straightforward manner,
        and $f$ is defined componentwise.
\end{itemize}
We see that this is indeed an effective symbolic dynamical system.
The projections $\pi_n:X \rightarrow X_n$ are effective maps as
well. Products are useful to build examples of systems with
particular properties, as illustrated in Propositions
\ref{soficuniversal} and \ref{propshadow_but_not_eff}.

\section{Finite automata} \label{sectautomata}

Consider an effective system $f: X \rightarrow X$ and two clopen
sets $U,V \subseteq X$. We would like to know if there is a point
of $U$ that eventually reaches $V$, that is, if there exists an
$x$ such that

\begin{equation} \label{firstorderhalting}
x\in U \,\textrm{and} \,\exists n\in\N : f^{n}(x) \in V.
\end{equation}

We  call \emph{halting problem of $f$}, the problem of answering
this question given $U$ and $V$. We will see later that this is
indeed a generalization of the halting problem traditionally
defined for Turing machines or counter machines.

\begin{figure}
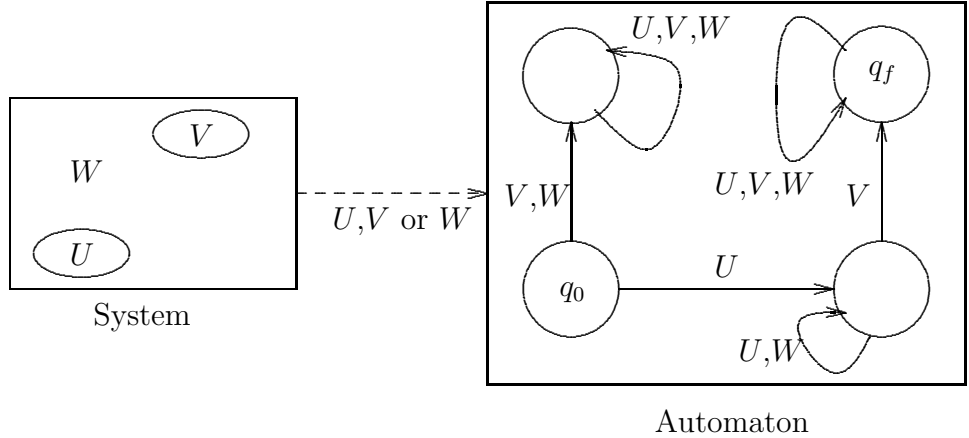

\begin{center}
\font\thinlinefont=cmr5
\begingroup\makeatletter\ifx\SetFigFont\undefined%
\gdef\SetFigFont#1#2#3#4#5{%
  \reset@font\fontsize{#1}{#2pt}%
  \fontfamily{#3}\fontseries{#4}\fontshape{#5}%
  \selectfont}%
\fi\endgroup%
\mbox{\beginpicture \setcoordinatesystem units
<1.00000cm,1.00000cm> \unitlength=1.00000cm \linethickness=1pt
\setplotsymbol ({\makebox(0,0)[l]{\tencirc\symbol{'160}}})
\setshadesymbol ({\thinlinefont .}) \setlinear
%
%
\linethickness= 0.500pt \setplotsymbol ({\thinlinefont .})
{\ellipticalarc axes ratio  0.635:0.318  360 degrees
    from  6.668 22.066 center at  6.032 22.066
}%
%
%
\linethickness= 0.500pt \setplotsymbol ({\thinlinefont .})
{\ellipticalarc axes ratio  0.635:0.318  360 degrees
    from  8.255 23.654 center at  7.620 23.654
}%
%
%
\linethickness= 0.500pt \setplotsymbol ({\thinlinefont .})
{\ellipticalarc axes ratio  0.635:0.635  360 degrees
    from 17.304 21.590 center at 16.669 21.590
}%
%
%
\linethickness= 0.500pt \setplotsymbol ({\thinlinefont .})
{\ellipticalarc axes ratio  0.635:0.635  360 degrees
    from 17.304 24.448 center at 16.669 24.448
}%
%
%
\linethickness= 0.500pt \setplotsymbol ({\thinlinefont .})
{\ellipticalarc axes ratio  0.635:0.635  360 degrees
    from 13.172 24.428 center at 12.537 24.428
}%
%
%
\linethickness= 0.500pt \setplotsymbol ({\thinlinefont .})
{\ellipticalarc axes ratio  0.635:0.635  360 degrees
    from 13.176 21.590 center at 12.541 21.590
}%
%
%
\linethickness= 0.500pt \setplotsymbol ({\thinlinefont .})
{\putrectangle corners at  5.080 24.130 and  8.890 21.590
}%
%
%
\linethickness= 0.500pt \setplotsymbol ({\thinlinefont .})
{\putrectangle corners at 11.430 25.400 and 17.780 20.320
}%
%
%
\linethickness= 0.500pt \setplotsymbol ({\thinlinefont .})
{\putrule from 13.176 21.590 to 16.034 21.590
%
%
\plot 15.780 21.526 16.034 21.590 15.780 21.654 /
}%
%
%
\linethickness= 0.500pt \setplotsymbol ({\thinlinefont .})
{\putrule from 16.669 22.225 to 16.669 23.812
%
%
\plot 16.732 23.559 16.669 23.812 16.605 23.559 /
}%
%
%
\linethickness= 0.500pt \setplotsymbol ({\thinlinefont .})
{\putrule from 12.541 22.225 to 12.541 23.812
%
%
\plot 12.605 23.559 12.541 23.812 12.478 23.559 /
}%
%
%
\linethickness= 0.500pt \setplotsymbol ({\thinlinefont .})
\setdashes < 0.1270cm> {\plot  8.890 22.860 11.430 22.860 /
\setsolid
%
%
\plot 11.176 22.796 11.430 22.860 11.176 22.924 /
\setdashes < 0.1270cm>
}%
%
%
\linethickness= 0.500pt \setplotsymbol ({\thinlinefont .})
\setsolid {\plot 12.859 23.971 12.865 23.965 / \plot 12.865 23.965
12.878 23.950 / \plot 12.878 23.950 12.901 23.927 / \plot 12.901
23.927 12.933 23.891 / \plot 12.933 23.891 12.977 23.848 / \plot
12.977 23.848 13.028 23.796 / \plot 13.028 23.796 13.083 23.741 /
\plot 13.083 23.741 13.142 23.686 / \plot 13.142 23.686 13.200
23.633 / \plot 13.200 23.633 13.257 23.586 / \plot 13.257 23.586
13.310 23.544 / \plot 13.310 23.544 13.360 23.510 / \plot 13.360
23.510 13.405 23.482 / \plot 13.405 23.482 13.447 23.461 / \plot
13.447 23.461 13.487 23.448 / \plot 13.487 23.448 13.523 23.444 /
\putrule from 13.523 23.444 to 13.559 23.444 \plot 13.559 23.444
13.593 23.453 / \plot 13.593 23.453 13.627 23.467 / \plot 13.627
23.467 13.652 23.487 / \plot 13.652 23.487 13.680 23.508 / \plot
13.680 23.508 13.708 23.533 / \plot 13.708 23.533 13.733 23.563 /
\plot 13.733 23.563 13.760 23.594 / \plot 13.760 23.594 13.786
23.633 / \plot 13.786 23.633 13.811 23.673 / \plot 13.811 23.673
13.837 23.715 / \plot 13.837 23.715 13.860 23.762 / \plot 13.860
23.762 13.881 23.810 / \plot 13.881 23.810 13.902 23.861 / \plot
13.902 23.861 13.921 23.914 / \plot 13.921 23.914 13.940 23.967 /
\plot 13.940 23.967 13.955 24.020 / \plot 13.955 24.020 13.968
24.073 / \plot 13.968 24.073 13.978 24.126 / \plot 13.978 24.126
13.987 24.177 / \plot 13.987 24.177 13.993 24.225 / \plot 13.993
24.225 13.995 24.274 / \putrule from 13.995 24.274 to 13.995
24.318 \plot 13.995 24.318 13.993 24.361 / \plot 13.993 24.361
13.989 24.401 / \plot 13.989 24.401 13.981 24.439 / \plot 13.981
24.439 13.970 24.473 / \plot 13.970 24.473 13.953 24.511 / \plot
13.953 24.511 13.932 24.547 / \plot 13.932 24.547 13.904 24.577 /
\plot 13.904 24.577 13.871 24.604 / \plot 13.871 24.604 13.830
24.627 / \plot 13.830 24.627 13.784 24.649 / \plot 13.784 24.649
13.727 24.668 / \plot 13.727 24.668 13.665 24.687 / \plot 13.665
24.687 13.593 24.701 / \plot 13.593 24.701 13.517 24.714 / \plot
13.517 24.714 13.434 24.727 / \plot 13.434 24.727 13.350 24.737 /
\plot 13.350 24.737 13.269 24.746 / \plot 13.269 24.746 13.195
24.752 / \plot 13.195 24.752 13.130 24.759 / \plot 13.130 24.759
13.081 24.761 / \plot 13.081 24.761 13.018 24.765 /
%
%
\plot 13.275 24.811 13.018 24.765 13.267 24.685 /
}%
%
%
\linethickness= 0.500pt \setplotsymbol ({\thinlinefont .}) {\plot
16.192 24.765 16.190 24.767 / \plot 16.190 24.767 16.184 24.773 /
\plot 16.184 24.773 16.176 24.784 / \plot 16.176 24.784 16.159
24.801 / \plot 16.159 24.801 16.140 24.824 / \plot 16.140 24.824
16.112 24.852 / \plot 16.112 24.852 16.080 24.884 / \plot 16.080
24.884 16.044 24.920 / \plot 16.044 24.920 16.006 24.960 / \plot
16.006 24.960 15.964 24.998 / \plot 15.964 24.998 15.919 25.038 /
\plot 15.919 25.038 15.875 25.076 / \plot 15.875 25.076 15.828
25.112 / \plot 15.828 25.112 15.782 25.142 / \plot 15.782 25.142
15.737 25.169 / \plot 15.737 25.169 15.691 25.190 / \plot 15.691
25.190 15.646 25.205 / \plot 15.646 25.205 15.602 25.212 / \plot
15.602 25.212 15.558 25.210 / \plot 15.558 25.210 15.515 25.197 /
\plot 15.515 25.197 15.473 25.174 / \plot 15.473 25.174 15.435
25.135 / \plot 15.435 25.135 15.399 25.082 / \plot 15.399 25.082
15.373 25.027 / \plot 15.373 25.027 15.350 24.964 / \plot 15.350
24.964 15.331 24.896 / \plot 15.331 24.896 15.316 24.824 / \plot
15.316 24.824 15.303 24.754 / \plot 15.303 24.754 15.293 24.682 /
\plot 15.293 24.682 15.287 24.613 / \plot 15.287 24.613 15.280
24.545 / \plot 15.280 24.545 15.276 24.477 / \plot 15.276 24.477
15.272 24.414 / \plot 15.272 24.414 15.270 24.350 / \plot 15.270
24.350 15.268 24.289 / \putrule from 15.268 24.289 to 15.268
24.229 \putrule from 15.268 24.229 to 15.268 24.170 \putrule from
15.268 24.170 to 15.268 24.111 \putrule from 15.268 24.111 to
15.268 24.050 \plot 15.268 24.050 15.270 23.990 / \plot 15.270
23.990 15.272 23.929 / \plot 15.272 23.929 15.276 23.868 / \plot
15.276 23.868 15.280 23.804 / \plot 15.280 23.804 15.287 23.738 /
\plot 15.287 23.738 15.293 23.673 / \plot 15.293 23.673 15.303
23.607 / \plot 15.303 23.607 15.316 23.544 / \plot 15.316 23.544
15.331 23.482 / \plot 15.331 23.482 15.350 23.425 / \plot 15.350
23.425 15.373 23.376 / \plot 15.373 23.376 15.399 23.336 / \plot
15.399 23.336 15.437 23.302 / \plot 15.437 23.302 15.481 23.290 /
\plot 15.481 23.290 15.526 23.294 / \plot 15.526 23.294 15.574
23.311 / \plot 15.574 23.311 15.623 23.343 / \plot 15.623 23.343
15.672 23.383 / \plot 15.672 23.383 15.720 23.434 / \plot 15.720
23.434 15.771 23.491 / \plot 15.771 23.491 15.820 23.554 / \plot
15.820 23.554 15.871 23.622 / \plot 15.871 23.622 15.919 23.694 /
\plot 15.919 23.694 15.968 23.766 / \plot 15.968 23.766 16.013
23.836 / \plot 16.013 23.836 16.055 23.904 / \plot 16.055 23.904
16.093 23.963 / \plot 16.093 23.963 16.125 24.016 / \plot 16.125
24.016 16.150 24.060 / \plot 16.150 24.060 16.192 24.130 /
%
%
\plot 16.115 23.880 16.192 24.130 16.007 23.946 /
}%
%
%
\linethickness= 0.500pt \setplotsymbol ({\thinlinefont .}) {\plot
16.510 20.955 16.508 20.951 / \plot 16.508 20.951 16.504 20.940 /
\plot 16.504 20.940 16.493 20.921 / \plot 16.493 20.921 16.480
20.896 / \plot 16.480 20.896 16.463 20.862 / \plot 16.463 20.862
16.442 20.822 / \plot 16.442 20.822 16.417 20.777 / \plot 16.417
20.777 16.391 20.733 / \plot 16.391 20.733 16.362 20.686 / \plot
16.362 20.686 16.330 20.642 / \plot 16.330 20.642 16.298 20.602 /
\plot 16.298 20.602 16.262 20.563 / \plot 16.262 20.563 16.224
20.532 / \plot 16.224 20.532 16.182 20.504 / \plot 16.182 20.504
16.135 20.485 / \plot 16.135 20.485 16.087 20.477 / \plot 16.087
20.477 16.034 20.479 / \plot 16.034 20.479 15.987 20.491 / \plot
15.987 20.491 15.941 20.513 / \plot 15.941 20.513 15.900 20.538 /
\plot 15.900 20.538 15.862 20.568 / \plot 15.862 20.568 15.828
20.597 / \plot 15.828 20.597 15.799 20.627 / \plot 15.799 20.627
15.771 20.657 / \plot 15.771 20.657 15.746 20.686 / \plot 15.746
20.686 15.725 20.714 / \plot 15.725 20.714 15.704 20.743 / \plot
15.704 20.743 15.682 20.773 / \plot 15.682 20.773 15.663 20.803 /
\plot 15.663 20.803 15.642 20.834 / \plot 15.642 20.834 15.623
20.868 / \plot 15.623 20.868 15.604 20.906 / \plot 15.604 20.906
15.587 20.944 / \plot 15.587 20.944 15.572 20.987 / \plot 15.572
20.987 15.560 21.029 / \plot 15.560 21.029 15.555 21.071 / \plot
15.555 21.071 15.558 21.114 / \plot 15.558 21.114 15.574 21.158 /
\plot 15.574 21.158 15.606 21.196 / \plot 15.606 21.196 15.646
21.224 / \plot 15.646 21.224 15.695 21.245 / \plot 15.695 21.245
15.748 21.258 / \plot 15.748 21.258 15.807 21.268 / \plot 15.807
21.268 15.867 21.275 / \plot 15.867 21.275 15.930 21.277 / \plot
15.930 21.277 15.991 21.279 / \putrule from 15.991 21.279 to
16.049 21.279 \plot 16.049 21.279 16.099 21.277 / \plot 16.099
21.277 16.140 21.275 / \plot 16.140 21.275 16.192 21.273 /
%
%
\plot 15.936 21.219 16.192 21.273 15.941 21.346 /
}%
%
%
\put{\SetFigFont{12}{14.4}{\rmdefault}{\mddefault}{\updefault}{$U$}%
} [lB] at  5.874 21.907
%
%
\put{\SetFigFont{12}{14.4}{\rmdefault}{\mddefault}{\updefault}{$V$}%
} [lB] at  7.461 23.495
%
%
\put{\SetFigFont{12}{14.4}{\rmdefault}{\mddefault}{\updefault}{$W$}%
} [lB] at  5.874 23.019
%
%
\put{\SetFigFont{12}{14.4}{\rmdefault}{\mddefault}{\updefault}{$U$}%
} [lB] at 14.446 21.749
%
%
\put{\SetFigFont{12}{14.4}{\rmdefault}{\mddefault}{\updefault}{$q_f$}%
} [lB] at 16.491 24.438
%
%
\put{\SetFigFont{12}{14.4}{\rmdefault}{\mddefault}{\updefault}{$V$,$W$}%
} [lB] at 11.648 22.701
%
%
\put{\SetFigFont{12}{14.4}{\rmdefault}{\mddefault}{\updefault}{$U$,$V$,$W$}%
} [lB] at 13.335 24.924
%
%
\put{\SetFigFont{12}{14.4}{\rmdefault}{\mddefault}{\updefault}{$V$}%
} [lB] at 16.192 22.701
%
%
\put{\SetFigFont{12}{14.4}{\rmdefault}{\mddefault}{\updefault}{$U$,$V$,$W$}%
} [lB] at 14.446 22.860
%
%
\put{\SetFigFont{12}{14.4}{\rmdefault}{\mddefault}{\updefault}{$q_0$}%
} [lB] at 12.384 21.500
%
%
\put{\SetFigFont{12}{14.4}{\rmdefault}{\mddefault}{\updefault}{$U$,$V$ or $W$}%
} [lB] at  9.366 22.384
%
%
\put{\SetFigFont{12}{14.4}{\rmdefault}{\mddefault}{\updefault}{System}%
} [lB] at  6.191 21.114
%
%
\put{\SetFigFont{12}{14.4}{\rmdefault}{\mddefault}{\updefault}{Automaton}%
} [lB] at 13.652 19.685
%
%
\put{\SetFigFont{12}{14.4}{\rmdefault}{\mddefault}{\updefault}{$U$,$W$}%
} [lB] at 14.764 20.637 \linethickness=0pt \putrectangle corners
at 5.055 25.425 and 17.805 19.685
\endpicture}
\end{center}
\caption{The symbolic system is partitioned into $U$, $V$ and
  $W= X \setminus (U \cup V)$. At every time step, the finite automaton is
  fed with the symbol $U$, $V$ or $W$ and jumps
  to a new state. It is possible to reach  the final state
  $q_f$ from the initial state $q_0$ iff it is possible that $q_f$ (and only $q_f$) is reached infinitely
  often from the initial state $q_0$ iff  there is a point of $U$ that eventually reaches $V$.
  Checking whether this is true given $U$ and $V$, is the
  \emph{halting problem} of $f$. The automaton can be considered as a finite automaton (the final state is $q_f$) or
  as a Muller automaton (for the family  $\{\{q_f\}\}$).}
    \label{fighalting1}
\end{figure}

Consider now another formulation of the halting problem. Suppose
that the system $f$ is only partially observable. All we can know
about $f$ is whether the system is currently in $U$, in $V$ or in
$W = X \setminus (U \cup V)$ (we suppose for simplicity that $U$
and $V$ are disjoint). The system is observed by a finite
automaton (formally defined below) as illustrated in Figure
\ref{fighalting1} . At every time step, the automaton jumps to a
new state, according to which set $U$, $V$ or $W$ the system is
currently in. The halting problem amounts to deciding whether it
is possible, for some initial point of the space $X$, that the
automaton eventually reaches the final state from the initial
state.

\begin{figure}[!h]
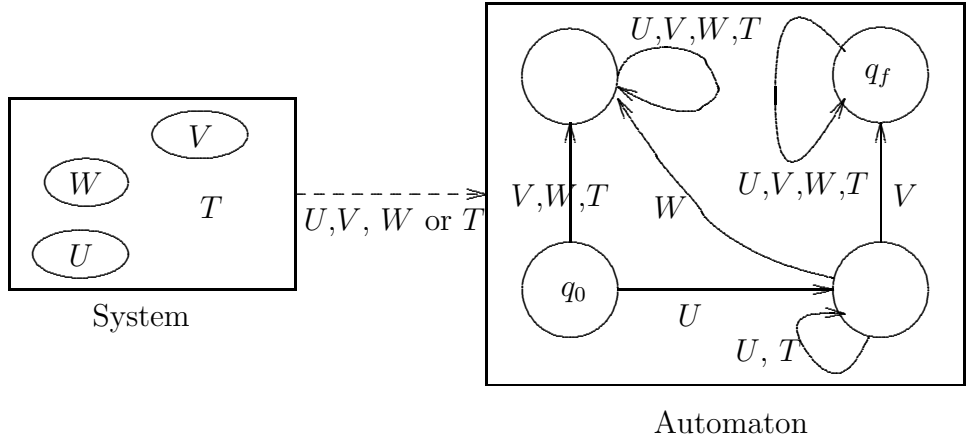

\begin{center}
\font\thinlinefont=cmr5
\begingroup\makeatletter\ifx\SetFigFont\undefined%
\gdef\SetFigFont#1#2#3#4#5{%
  \reset@font\fontsize{#1}{#2pt}%
  \fontfamily{#3}\fontseries{#4}\fontshape{#5}%
  \selectfont}%
\fi\endgroup%
\mbox{\beginpicture \setcoordinatesystem units
<1.00000cm,1.00000cm> \unitlength=1.00000cm \linethickness=1pt
\setplotsymbol ({\makebox(0,0)[l]{\tencirc\symbol{'160}}})
\setshadesymbol ({\thinlinefont .}) \setlinear
%
%
\linethickness= 0.500pt \setplotsymbol ({\thinlinefont .})
{\ellipticalarc axes ratio  0.635:0.318  360 degrees
    from  6.668 22.066 center at  6.032 22.066
}%
%
%
\linethickness= 0.500pt \setplotsymbol ({\thinlinefont .})
{\ellipticalarc axes ratio  0.635:0.318  360 degrees
    from  8.255 23.654 center at  7.620 23.654
}%
%
%
\linethickness= 0.500pt \setplotsymbol ({\thinlinefont .})
{\ellipticalarc axes ratio  0.635:0.635  360 degrees
    from 17.304 21.590 center at 16.669 21.590
}%
%
%
\linethickness= 0.500pt \setplotsymbol ({\thinlinefont .})
{\ellipticalarc axes ratio  0.635:0.635  360 degrees
    from 17.304 24.448 center at 16.669 24.448
}%
%
%
\linethickness= 0.500pt \setplotsymbol ({\thinlinefont .})
{\ellipticalarc axes ratio  0.635:0.635  360 degrees
    from 13.176 21.590 center at 12.541 21.590
}%
%
%
\linethickness= 0.500pt \setplotsymbol ({\thinlinefont .})
{\ellipticalarc axes ratio  0.635:0.635  360 degrees
    from 13.172 24.428 center at 12.537 24.428
}%
%
%
\linethickness= 0.500pt \setplotsymbol ({\thinlinefont .})
{\ellipticalarc axes ratio  0.555:0.318  360 degrees
    from  6.665 23.019 center at  6.111 23.019
}%
%
%
\linethickness= 0.500pt \setplotsymbol ({\thinlinefont .})
{\putrectangle corners at  5.080 24.130 and  8.890 21.590
}%
%
%
\linethickness= 0.500pt \setplotsymbol ({\thinlinefont .})
{\putrectangle corners at 11.430 25.400 and 17.780 20.320
}%
%
%
\linethickness= 0.500pt \setplotsymbol ({\thinlinefont .})
{\putrule from 13.176 21.590 to 16.034 21.590
%
%
\plot 15.780 21.526 16.034 21.590 15.780 21.654 /
}%
%
%
\linethickness= 0.500pt \setplotsymbol ({\thinlinefont .})
{\putrule from 16.669 22.225 to 16.669 23.812
%
%
\plot 16.732 23.559 16.669 23.812 16.605 23.559 /
}%
%
%
\linethickness= 0.500pt \setplotsymbol ({\thinlinefont .})
{\putrule from 12.541 22.225 to 12.541 23.812
%
%
\plot 12.605 23.559 12.541 23.812 12.478 23.559 /
}%
%
%
\linethickness= 0.500pt \setplotsymbol ({\thinlinefont .})
\setdashes < 0.1270cm> {\plot  8.890 22.860 11.430 22.860 /
\setsolid
%
%
\plot 11.176 22.796 11.430 22.860 11.176 22.924 /
\setdashes < 0.1270cm>
}%
%
%
\linethickness= 0.500pt \setplotsymbol ({\thinlinefont .})
\setsolid {\plot 16.192 24.765 16.190 24.767 / \plot 16.190 24.767
16.184 24.773 / \plot 16.184 24.773 16.176 24.784 / \plot 16.176
24.784 16.159 24.801 / \plot 16.159 24.801 16.140 24.824 / \plot
16.140 24.824 16.112 24.852 / \plot 16.112 24.852 16.080 24.884 /
\plot 16.080 24.884 16.044 24.920 / \plot 16.044 24.920 16.006
24.960 / \plot 16.006 24.960 15.964 24.998 / \plot 15.964 24.998
15.919 25.038 / \plot 15.919 25.038 15.875 25.076 / \plot 15.875
25.076 15.828 25.112 / \plot 15.828 25.112 15.782 25.142 / \plot
15.782 25.142 15.737 25.169 / \plot 15.737 25.169 15.691 25.190 /
\plot 15.691 25.190 15.646 25.205 / \plot 15.646 25.205 15.602
25.212 / \plot 15.602 25.212 15.558 25.210 / \plot 15.558 25.210
15.515 25.197 / \plot 15.515 25.197 15.473 25.174 / \plot 15.473
25.174 15.435 25.135 / \plot 15.435 25.135 15.399 25.082 / \plot
15.399 25.082 15.373 25.027 / \plot 15.373 25.027 15.350 24.964 /
\plot 15.350 24.964 15.331 24.896 / \plot 15.331 24.896 15.316
24.824 / \plot 15.316 24.824 15.303 24.754 / \plot 15.303 24.754
15.293 24.682 / \plot 15.293 24.682 15.287 24.613 / \plot 15.287
24.613 15.280 24.545 / \plot 15.280 24.545 15.276 24.477 / \plot
15.276 24.477 15.272 24.414 / \plot 15.272 24.414 15.270 24.350 /
\plot 15.270 24.350 15.268 24.289 / \putrule from 15.268 24.289 to
15.268 24.229 \putrule from 15.268 24.229 to 15.268 24.170
\putrule from 15.268 24.170 to 15.268 24.111 \putrule from 15.268
24.111 to 15.268 24.050 \plot 15.268 24.050 15.270 23.990 / \plot
15.270 23.990 15.272 23.929 / \plot 15.272 23.929 15.276 23.868 /
\plot 15.276 23.868 15.280 23.804 / \plot 15.280 23.804 15.287
23.738 / \plot 15.287 23.738 15.293 23.673 / \plot 15.293 23.673
15.303 23.607 / \plot 15.303 23.607 15.316 23.544 / \plot 15.316
23.544 15.331 23.482 / \plot 15.331 23.482 15.350 23.425 / \plot
15.350 23.425 15.373 23.376 / \plot 15.373 23.376 15.399 23.336 /
\plot 15.399 23.336 15.437 23.302 / \plot 15.437 23.302 15.481
23.290 / \plot 15.481 23.290 15.526 23.294 / \plot 15.526 23.294
15.574 23.311 / \plot 15.574 23.311 15.623 23.343 / \plot 15.623
23.343 15.672 23.383 / \plot 15.672 23.383 15.720 23.434 / \plot
15.720 23.434 15.771 23.491 / \plot 15.771 23.491 15.820 23.554 /
\plot 15.820 23.554 15.871 23.622 / \plot 15.871 23.622 15.919
23.694 / \plot 15.919 23.694 15.968 23.766 / \plot 15.968 23.766
16.013 23.836 / \plot 16.013 23.836 16.055 23.904 / \plot 16.055
23.904 16.093 23.963 / \plot 16.093 23.963 16.125 24.016 / \plot
16.125 24.016 16.150 24.060 / \plot 16.150 24.060 16.192 24.130 /
%
%
\plot 16.115 23.880 16.192 24.130 16.007 23.946 /
}%
%
%
\linethickness= 0.500pt \setplotsymbol ({\thinlinefont .}) {\plot
16.510 20.955 16.508 20.951 / \plot 16.508 20.951 16.504 20.940 /
\plot 16.504 20.940 16.493 20.921 / \plot 16.493 20.921 16.480
20.896 / \plot 16.480 20.896 16.463 20.862 / \plot 16.463 20.862
16.442 20.822 / \plot 16.442 20.822 16.417 20.777 / \plot 16.417
20.777 16.391 20.733 / \plot 16.391 20.733 16.362 20.686 / \plot
16.362 20.686 16.330 20.642 / \plot 16.330 20.642 16.298 20.602 /
\plot 16.298 20.602 16.262 20.563 / \plot 16.262 20.563 16.224
20.532 / \plot 16.224 20.532 16.182 20.504 / \plot 16.182 20.504
16.135 20.485 / \plot 16.135 20.485 16.087 20.477 / \plot 16.087
20.477 16.034 20.479 / \plot 16.034 20.479 15.987 20.491 / \plot
15.987 20.491 15.941 20.513 / \plot 15.941 20.513 15.900 20.538 /
\plot 15.900 20.538 15.862 20.568 / \plot 15.862 20.568 15.828
20.597 / \plot 15.828 20.597 15.799 20.627 / \plot 15.799 20.627
15.771 20.657 / \plot 15.771 20.657 15.746 20.686 / \plot 15.746
20.686 15.725 20.714 / \plot 15.725 20.714 15.704 20.743 / \plot
15.704 20.743 15.682 20.773 / \plot 15.682 20.773 15.663 20.803 /
\plot 15.663 20.803 15.642 20.834 / \plot 15.642 20.834 15.623
20.868 / \plot 15.623 20.868 15.604 20.906 / \plot 15.604 20.906
15.587 20.944 / \plot 15.587 20.944 15.572 20.987 / \plot 15.572
20.987 15.560 21.029 / \plot 15.560 21.029 15.555 21.071 / \plot
15.555 21.071 15.558 21.114 / \plot 15.558 21.114 15.574 21.158 /
\plot 15.574 21.158 15.606 21.196 / \plot 15.606 21.196 15.646
21.224 / \plot 15.646 21.224 15.695 21.245 / \plot 15.695 21.245
15.748 21.258 / \plot 15.748 21.258 15.807 21.268 / \plot 15.807
21.268 15.867 21.275 / \plot 15.867 21.275 15.930 21.277 / \plot
15.930 21.277 15.991 21.279 / \putrule from 15.991 21.279 to
16.049 21.279 \plot 16.049 21.279 16.099 21.277 / \plot 16.099
21.277 16.140 21.275 / \plot 16.140 21.275 16.192 21.273 /
%
%
\plot 15.936 21.219 16.192 21.273 15.941 21.346 /
}%
%
%
\linethickness= 0.500pt \setplotsymbol ({\thinlinefont .})
{\putrule from 16.034 21.749 to 16.032 21.749 \plot 16.032 21.749
16.023 21.751 / \plot 16.023 21.751 16.010 21.753 / \plot 16.010
21.753 15.991 21.757 / \plot 15.991 21.757 15.964 21.764 / \plot
15.964 21.764 15.930 21.772 / \plot 15.930 21.772 15.890 21.783 /
\plot 15.890 21.783 15.841 21.793 / \plot 15.841 21.793 15.788
21.806 / \plot 15.788 21.806 15.731 21.821 / \plot 15.731 21.821
15.672 21.838 / \plot 15.672 21.838 15.608 21.857 / \plot 15.608
21.857 15.543 21.876 / \plot 15.543 21.876 15.475 21.897 / \plot
15.475 21.897 15.407 21.920 / \plot 15.407 21.920 15.335 21.946 /
\plot 15.335 21.946 15.261 21.975 / \plot 15.261 21.975 15.185
22.007 / \plot 15.185 22.007 15.107 22.043 / \plot 15.107 22.043
15.024 22.081 / \plot 15.024 22.081 14.937 22.126 / \plot 14.937
22.126 14.851 22.174 / \plot 14.851 22.174 14.764 22.225 / \plot
14.764 22.225 14.673 22.284 / \plot 14.673 22.284 14.590 22.339 /
\plot 14.590 22.339 14.518 22.392 / \plot 14.518 22.392 14.459
22.439 / \plot 14.459 22.439 14.412 22.479 / \plot 14.412 22.479
14.374 22.511 / \plot 14.374 22.511 14.349 22.538 / \plot 14.349
22.538 14.330 22.562 / \plot 14.330 22.562 14.317 22.578 / \plot
14.317 22.578 14.307 22.595 / \plot 14.307 22.595 14.300 22.608 /
\plot 14.300 22.608 14.294 22.623 / \plot 14.294 22.623 14.287
22.638 / \plot 14.287 22.638 14.277 22.657 / \plot 14.277 22.657
14.262 22.680 / \plot 14.262 22.680 14.241 22.708 / \plot 14.241
22.708 14.213 22.744 / \plot 14.213 22.744 14.180 22.786 / \plot
14.180 22.786 14.135 22.835 / \plot 14.135 22.835 14.084 22.892 /
\plot 14.084 22.892 14.027 22.953 / \plot 14.027 22.953 13.970
23.019 / \plot 13.970 23.019 13.902 23.095 / \plot 13.902 23.095
13.843 23.165 / \plot 13.843 23.165 13.794 23.224 / \plot 13.794
23.224 13.758 23.269 / \plot 13.758 23.269 13.731 23.302 / \plot
13.731 23.302 13.714 23.328 / \plot 13.714 23.328 13.703 23.343 /
\plot 13.703 23.343 13.697 23.355 / \plot 13.697 23.355 13.693
23.364 / \plot 13.693 23.364 13.688 23.370 / \plot 13.688 23.370
13.684 23.381 / \plot 13.684 23.381 13.676 23.396 / \plot 13.676
23.396 13.661 23.417 / \plot 13.661 23.417 13.642 23.446 / \plot
13.642 23.446 13.614 23.487 / \plot 13.614 23.487 13.578 23.535 /
\plot 13.578 23.535 13.538 23.592 / \plot 13.538 23.592 13.494
23.654 / \plot 13.494 23.654 13.445 23.721 / \plot 13.445 23.721
13.403 23.783 / \plot 13.403 23.783 13.367 23.836 / \plot 13.367
23.836 13.335 23.882 / \plot 13.335 23.882 13.305 23.925 / \plot
13.305 23.925 13.282 23.963 / \plot 13.282 23.963 13.261 23.995 /
\plot 13.261 23.995 13.240 24.026 / \plot 13.240 24.026 13.223
24.054 / \plot 13.223 24.054 13.208 24.077 / \plot 13.208 24.077
13.197 24.096 / \plot 13.197 24.096 13.176 24.130 /
%
%
\plot 13.365 23.948 13.176 24.130 13.257 23.881 /
}%
%
%
\linethickness= 0.500pt \setplotsymbol ({\thinlinefont .}) {\plot
13.176 24.448 13.178 24.452 / \plot 13.178 24.452 13.180 24.458 /
\plot 13.180 24.458 13.185 24.469 / \plot 13.185 24.469 13.191
24.486 / \plot 13.191 24.486 13.200 24.507 / \plot 13.200 24.507
13.212 24.530 / \plot 13.212 24.530 13.225 24.558 / \plot 13.225
24.558 13.242 24.583 / \plot 13.242 24.583 13.263 24.610 / \plot
13.263 24.610 13.286 24.638 / \plot 13.286 24.638 13.314 24.666 /
\plot 13.314 24.666 13.348 24.693 / \plot 13.348 24.693 13.388
24.718 / \plot 13.388 24.718 13.437 24.744 / \plot 13.437 24.744
13.494 24.765 / \plot 13.494 24.765 13.555 24.782 / \plot 13.555
24.782 13.612 24.795 / \plot 13.612 24.795 13.663 24.805 / \plot
13.663 24.805 13.705 24.809 / \plot 13.705 24.809 13.741 24.814 /
\plot 13.741 24.814 13.769 24.816 / \plot 13.769 24.816 13.790
24.818 / \putrule from 13.790 24.818 to 13.811 24.818 \putrule
from 13.811 24.818 to 13.832 24.818 \plot 13.832 24.818 13.854
24.816 / \plot 13.854 24.816 13.881 24.814 / \plot 13.881 24.814
13.917 24.809 / \plot 13.917 24.809 13.959 24.805 / \plot 13.959
24.805 14.010 24.795 / \plot 14.010 24.795 14.067 24.782 / \plot
14.067 24.782 14.129 24.765 / \plot 14.129 24.765 14.186 24.744 /
\plot 14.186 24.744 14.232 24.721 / \plot 14.232 24.721 14.271
24.699 / \plot 14.271 24.699 14.296 24.680 / \plot 14.296 24.680
14.315 24.666 / \plot 14.315 24.666 14.328 24.653 / \plot 14.328
24.653 14.336 24.642 / \plot 14.336 24.642 14.340 24.632 / \plot
14.340 24.632 14.345 24.623 / \plot 14.345 24.623 14.351 24.610 /
\plot 14.351 24.610 14.357 24.596 / \plot 14.357 24.596 14.368
24.577 / \plot 14.368 24.577 14.383 24.551 / \plot 14.383 24.551
14.404 24.522 / \plot 14.404 24.522 14.425 24.486 / \plot 14.425
24.486 14.446 24.448 / \plot 14.446 24.448 14.463 24.409 / \plot
14.463 24.409 14.478 24.373 / \plot 14.478 24.373 14.486 24.344 /
\plot 14.486 24.344 14.495 24.323 / \plot 14.495 24.323 14.501
24.304 / \plot 14.501 24.304 14.506 24.293 / \plot 14.506 24.293
14.510 24.282 / \plot 14.510 24.282 14.512 24.276 / \plot 14.512
24.276 14.514 24.268 / \plot 14.514 24.268 14.516 24.259 / \plot
14.516 24.259 14.514 24.249 / \plot 14.514 24.249 14.512 24.234 /
\plot 14.512 24.234 14.503 24.213 / \plot 14.503 24.213 14.491
24.189 / \plot 14.491 24.189 14.472 24.160 / \plot 14.472 24.160
14.446 24.130 / \plot 14.446 24.130 14.415 24.102 / \plot 14.415
24.102 14.381 24.077 / \plot 14.381 24.077 14.351 24.056 / \plot
14.351 24.056 14.328 24.039 / \plot 14.328 24.039 14.309 24.026 /
\plot 14.309 24.026 14.294 24.016 / \plot 14.294 24.016 14.283
24.005 / \plot 14.283 24.005 14.275 23.997 / \plot 14.275 23.997
14.264 23.990 / \plot 14.264 23.990 14.249 23.984 / \plot 14.249
23.984 14.230 23.978 / \plot 14.230 23.978 14.203 23.971 / \plot
14.203 23.971 14.163 23.967 / \plot 14.163 23.967 14.112 23.963 /
\plot 14.112 23.963 14.046 23.965 / \plot 14.046 23.965 13.970
23.971 / \plot 13.970 23.971 13.906 23.982 / \plot 13.906 23.982
13.841 23.995 / \plot 13.841 23.995 13.780 24.011 / \plot 13.780
24.011 13.718 24.031 / \plot 13.718 24.031 13.661 24.050 / \plot
13.661 24.050 13.606 24.071 / \plot 13.606 24.071 13.553 24.094 /
\plot 13.553 24.094 13.502 24.117 / \plot 13.502 24.117 13.451
24.141 / \plot 13.451 24.141 13.405 24.164 / \plot 13.405 24.164
13.358 24.187 / \plot 13.358 24.187 13.316 24.210 / \plot 13.316
24.210 13.280 24.232 / \plot 13.280 24.232 13.246 24.249 / \plot
13.246 24.249 13.221 24.263 / \plot 13.221 24.263 13.176 24.289 /
%
%
\plot 13.428 24.218 13.176 24.289 13.365 24.108 /
}%
%
%
\put{\SetFigFont{12}{14.4}{\rmdefault}{\mddefault}{\updefault}{$U$}%
} [lB] at  5.874 21.907
%
%
\put{\SetFigFont{12}{14.4}{\rmdefault}{\mddefault}{\updefault}{$V$}%
} [lB] at  7.461 23.495
%
%
\put{\SetFigFont{12}{14.4}{\rmdefault}{\mddefault}{\updefault}{$q_f$}%
} [lB] at 16.451 24.408
%
%
\put{\SetFigFont{12}{14.4}{\rmdefault}{\mddefault}{\updefault}{$V$,$W$,$T$}%
} [lB] at 11.748 22.701
%
%
\put{\SetFigFont{12}{14.4}{\rmdefault}{\mddefault}{\updefault}{$U$,$V$,$W$,$T$}%
} [lB] at 13.335 24.924
%
%
\put{\SetFigFont{12}{14.4}{\rmdefault}{\mddefault}{\updefault}{$q_0$}%
} [lB] at 12.414 21.500
%
%
\put{\SetFigFont{12}{14.4}{\rmdefault}{\mddefault}{\updefault}{$U$,$V$, $W$ or $T$}%
} [lB] at  8.996 22.384
%
%
\put{\SetFigFont{12}{14.4}{\rmdefault}{\mddefault}{\updefault}{System}%
} [lB] at  6.191 21.114
%
%
\put{\SetFigFont{12}{14.4}{\rmdefault}{\mddefault}{\updefault}{Automaton}%
} [lB] at 13.652 19.685
%
%
\put{\SetFigFont{12}{14.4}{\rmdefault}{\mddefault}{\updefault}{$U$, $T$}%
} [lB] at 14.740 20.637
%
%
\put{\SetFigFont{12}{14.4}{\rmdefault}{\mddefault}{\updefault}{$U$}%
} [lB] at 13.970 21.114
%
%
\put{\SetFigFont{12}{14.4}{\rmdefault}{\mddefault}{\updefault}{$W$}%
} [lB] at 13.652 22.543
%
%
\put{\SetFigFont{12}{14.4}{\rmdefault}{\mddefault}{\updefault}{$U$,$V$,$W$,$T$}%
} [lB] at 14.764 22.860
%
%
\put{\SetFigFont{12}{14.4}{\rmdefault}{\mddefault}{\updefault}{$V$}%
} [lB] at 16.828 22.701
%
%
\put{\SetFigFont{12}{14.4}{\rmdefault}{\mddefault}{\updefault}{$W$}%
} [lB] at  5.874 22.860
%
%
\put{\SetFigFont{12}{14.4}{\rmdefault}{\mddefault}{\updefault}{$T$}%
} [lB] at  7.620 22.543 \linethickness=0pt \putrectangle corners
at 5.055 25.425 and 17.805 19.685
\endpicture}
\end{center}
 \caption{The symbolic system is partitioned into $U$, $V$, $W$ and
  $T=X \setminus (U \cup V \cup W)$.   There is a point of $U$ that stays in $X \setminus W$ until it eventually
  reaches $V$, iff it is possible that $q_f$ (and only $q_f$) is reached infinitely often from the initial state  $q_0$.}\label{fighalting2}
\end{figure}

We would like also consider variants of the halting problem. For
instance, given three disjoint clopen sets $U$, $V$ and $W$, we
want to check whether the following formula is satisfied for some
$x$:

\begin{equation} \label{firstordervariant1}
x \in U \, \textrm{and} \, \exists n : f^n(x) \in V \,
\textrm{and} \, \forall m < n : f^m(x) \notin W,
\end{equation}
where $n$ and $m$ are non-negative integers. A finite automaton
which accepts exactly points with this property is constructed in
Figure \ref{fighalting2}.

\begin{figure}
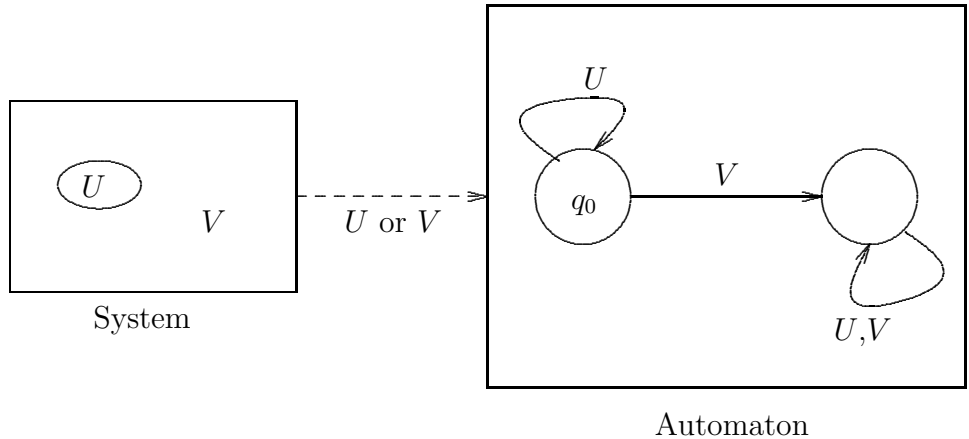

\begin{center}
\font\thinlinefont=cmr5
\begingroup\makeatletter\ifx\SetFigFont\undefined%
\gdef\SetFigFont#1#2#3#4#5{%
  \reset@font\fontsize{#1}{#2pt}%
  \fontfamily{#3}\fontseries{#4}\fontshape{#5}%
  \selectfont}%
\fi\endgroup%
\mbox{\beginpicture \setcoordinatesystem units
<1.00000cm,1.00000cm> \unitlength=1.00000cm \linethickness=1pt
\setplotsymbol ({\makebox(0,0)[l]{\tencirc\symbol{'160}}})
\setshadesymbol ({\thinlinefont .}) \setlinear
%
%
\linethickness= 0.500pt \setplotsymbol ({\thinlinefont .})
{\ellipticalarc axes ratio  0.555:0.318  360 degrees
    from  6.824 23.019 center at  6.270 23.019
}%
%
%
\linethickness= 0.500pt \setplotsymbol ({\thinlinefont .})
{\ellipticalarc axes ratio  0.635:0.635  360 degrees
    from 17.145 22.860 center at 16.510 22.860
}%
%
%
\linethickness= 0.500pt \setplotsymbol ({\thinlinefont .})
{\ellipticalarc axes ratio  0.635:0.635  360 degrees
    from 13.335 22.860 center at 12.700 22.860
}%
%
%
\linethickness= 0.500pt \setplotsymbol ({\thinlinefont .})
{\putrectangle corners at  5.080 24.130 and  8.890 21.590
}%
%
%
\linethickness= 0.500pt \setplotsymbol ({\thinlinefont .})
{\putrectangle corners at 11.430 25.400 and 17.780 20.320
}%
%
%
\linethickness= 0.500pt \setplotsymbol ({\thinlinefont .})
\setdashes < 0.1270cm> {\plot  8.890 22.860 11.430 22.860 /
\setsolid
%
%
\plot 11.176 22.796 11.430 22.860 11.176 22.924 /
\setdashes < 0.1270cm>
}%
%
%
\linethickness= 0.500pt \setplotsymbol ({\thinlinefont .})
\setsolid {\putrule from 13.335 22.860 to 15.875 22.860
%
%
\plot 15.621 22.796 15.875 22.860 15.621 22.924 /
}%
%
%
\linethickness= 0.500pt \setplotsymbol ({\thinlinefont .}) {\plot
16.986 22.384 16.990 22.382 / \plot 16.990 22.382 17.001 22.375 /
\plot 17.001 22.375 17.018 22.365 / \plot 17.018 22.365 17.041
22.348 / \plot 17.041 22.348 17.075 22.327 / \plot 17.075 22.327
17.113 22.299 / \plot 17.113 22.299 17.158 22.267 / \plot 17.158
22.267 17.206 22.233 / \plot 17.206 22.233 17.255 22.195 / \plot
17.255 22.195 17.302 22.155 / \plot 17.302 22.155 17.346 22.115 /
\plot 17.346 22.115 17.388 22.075 / \plot 17.388 22.075 17.424
22.032 / \plot 17.424 22.032 17.456 21.988 / \plot 17.456 21.988
17.477 21.941 / \plot 17.477 21.941 17.492 21.895 / \plot 17.492
21.895 17.496 21.846 / \plot 17.496 21.846 17.486 21.797 / \plot
17.486 21.797 17.462 21.749 / \plot 17.462 21.749 17.429 21.706 /
\plot 17.429 21.706 17.384 21.668 / \plot 17.384 21.668 17.335
21.634 / \plot 17.335 21.634 17.283 21.605 / \plot 17.283 21.605
17.230 21.577 / \plot 17.230 21.577 17.177 21.554 / \plot 17.177
21.554 17.126 21.533 / \plot 17.126 21.533 17.077 21.516 / \plot
17.077 21.516 17.029 21.499 / \plot 17.029 21.499 16.980 21.484 /
\plot 16.980 21.484 16.933 21.471 / \plot 16.933 21.471 16.887
21.459 / \plot 16.887 21.459 16.838 21.446 / \plot 16.838 21.446
16.787 21.435 / \plot 16.787 21.435 16.736 21.425 / \plot 16.736
21.425 16.684 21.414 / \plot 16.684 21.414 16.626 21.408 / \plot
16.626 21.408 16.569 21.402 / \putrule from 16.569 21.402 to
16.512 21.402 \plot 16.512 21.402 16.453 21.404 / \plot 16.453
21.404 16.400 21.414 / \plot 16.400 21.414 16.351 21.431 / \plot
16.351 21.431 16.311 21.461 / \plot 16.311 21.461 16.281 21.497 /
\plot 16.281 21.497 16.264 21.541 / \plot 16.264 21.541 16.258
21.588 / \plot 16.258 21.588 16.260 21.639 / \plot 16.260 21.639
16.269 21.692 / \plot 16.269 21.692 16.284 21.749 / \plot 16.284
21.749 16.303 21.806 / \plot 16.303 21.806 16.326 21.863 / \plot
16.326 21.863 16.351 21.922 / \plot 16.351 21.922 16.379 21.979 /
\plot 16.379 21.979 16.406 22.035 / \plot 16.406 22.035 16.432
22.085 / \plot 16.432 22.085 16.455 22.128 / \plot 16.455 22.128
16.476 22.164 / \plot 16.476 22.164 16.510 22.225 /
%
%
\plot 16.443 21.972 16.510 22.225 16.332 22.033 /
}%
%
%
\linethickness= 0.500pt \setplotsymbol ({\thinlinefont .}) {\plot
12.383 23.336 12.378 23.338 / \plot 12.378 23.338 12.368 23.347 /
\plot 12.368 23.347 12.349 23.357 / \plot 12.349 23.357 12.323
23.376 / \plot 12.323 23.376 12.289 23.400 / \plot 12.289 23.400
12.247 23.429 / \plot 12.247 23.429 12.200 23.463 / \plot 12.200
23.463 12.150 23.499 / \plot 12.150 23.499 12.099 23.539 / \plot
12.099 23.539 12.050 23.582 / \plot 12.050 23.582 12.004 23.624 /
\plot 12.004 23.624 11.961 23.666 / \plot 11.961 23.666 11.923
23.709 / \plot 11.923 23.709 11.894 23.753 / \plot 11.894 23.753
11.872 23.798 / \plot 11.872 23.798 11.860 23.842 / \putrule from
11.860 23.842 to 11.860 23.887 \plot 11.860 23.887 11.874 23.929 /
\plot 11.874 23.929 11.906 23.971 / \plot 11.906 23.971 11.944
24.003 / \plot 11.944 24.003 11.991 24.031 / \plot 11.991 24.031
12.044 24.054 / \plot 12.044 24.054 12.101 24.075 / \plot 12.101
24.075 12.156 24.092 / \plot 12.156 24.092 12.213 24.107 / \plot
12.213 24.107 12.268 24.119 / \plot 12.268 24.119 12.323 24.128 /
\plot 12.323 24.128 12.376 24.136 / \plot 12.376 24.136 12.427
24.145 / \plot 12.427 24.145 12.478 24.151 / \plot 12.478 24.151
12.529 24.158 / \plot 12.529 24.158 12.579 24.162 / \plot 12.579
24.162 12.630 24.166 / \plot 12.630 24.166 12.681 24.170 / \plot
12.681 24.170 12.736 24.172 / \plot 12.736 24.172 12.791 24.174 /
\putrule from 12.791 24.174 to 12.846 24.174 \putrule from 12.846
24.174 to 12.905 24.174 \plot 12.905 24.174 12.965 24.172 / \plot
12.965 24.172 13.024 24.168 / \plot 13.024 24.168 13.079 24.160 /
\plot 13.079 24.160 13.132 24.147 / \plot 13.132 24.147 13.176
24.130 / \plot 13.176 24.130 13.216 24.102 / \plot 13.216 24.102
13.240 24.069 / \plot 13.240 24.069 13.250 24.033 / \putrule from
13.250 24.033 to 13.250 23.992 \plot 13.250 23.992 13.240 23.952 /
\plot 13.240 23.952 13.219 23.910 / \plot 13.219 23.910 13.193
23.865 / \plot 13.193 23.865 13.161 23.819 / \plot 13.161 23.819
13.125 23.774 / \plot 13.125 23.774 13.087 23.728 / \plot 13.087
23.728 13.047 23.683 / \plot 13.047 23.683 13.007 23.641 / \plot
13.007 23.641 12.969 23.603 / \plot 12.969 23.603 12.935 23.569 /
\plot 12.935 23.569 12.907 23.542 / \plot 12.907 23.542 12.859
23.495 /
%
%
\plot 12.998 23.716 12.859 23.495 13.086 23.625 /
}%
%
%
\put{\SetFigFont{12}{14.4}{\rmdefault}{\mddefault}{\updefault}{System}%
} [lB] at  6.191 21.114
%
%
\put{\SetFigFont{12}{14.4}{\rmdefault}{\mddefault}{\updefault}{Automaton}%
} [lB] at 13.652 19.685
%
%
\put{\SetFigFont{12}{14.4}{\rmdefault}{\mddefault}{\updefault}{$U$}%
} [lB] at  6.032 22.860
%
%
\put{\SetFigFont{12}{14.4}{\rmdefault}{\mddefault}{\updefault}{$V$}%
} [lB] at  7.620 22.384
%
%
\put{\SetFigFont{12}{14.4}{\rmdefault}{\mddefault}{\updefault}{$U$,$V$}%
} [lB] at 16.034 20.955
%
%
\put{\SetFigFont{12}{14.4}{\rmdefault}{\mddefault}{\updefault}{$V$}%
} [lB] at 14.446 23.019
%
%
\put{\SetFigFont{12}{14.4}{\rmdefault}{\mddefault}{\updefault}{$U$}%
} [lB] at 12.700 24.289
%
%
\put{\SetFigFont{12}{14.4}{\rmdefault}{\mddefault}{\updefault}{$q_0$}%
} [lB] at 12.543 22.701
%
%
\put{\SetFigFont{12}{14.4}{\rmdefault}{\mddefault}{\updefault}{$U$ or $V$}%
} [lB] at  9.525 22.384 \linethickness=0pt \putrectangle corners
at  5.055 25.425 and 17.805 19.685
\endpicture}
\end{center}
  \caption{The system is partitioned into $U$ and $V$. There is a point that never leaves $U$ iff
  it is possible that $q_0$ (and only $q_0$) is reached infinitely often from the initial state $q_0$.}\label{fighalting3}
\end{figure}

We can also ask
whether the formula
\begin{equation} \label{firstordervariant2}
 \forall n :  f^n(x) \in U \
\end{equation}
is satisfied for some $x \in X$. This is the case if and only if
the automaton in Figure \ref{fighalting3}, starting from the
initial state and observing the system $f$, reaches infinitely
often the final state from the initial state. This leads us to the
theory of $\omega$-regular languages which can be recognized by
Muller or B\"uchi automata.

In general we are interested in all properties that can be
observed by automata. A (deterministic) \emph{finite automaton} is
given by a finite \emph{set of states} $Q$, an \emph{initial
state} $q_0\in Q$, a set of \emph{final} states $Q_1 \subseteq Q$,
a finite \emph{input alphabet} $A$ and a transition function
$\Delta: Q \times A \rightarrow Q$. The transition function is
extended to $\Delta: Q\times A^* \to Q$ by $\Delta(q,ua) =
\Delta(\Delta(q,u),a)$. A language  $L\subseteq A^*$ is
\emph{regular} if there exists a finite automaton which accepts
$L$, i.e., $u\in L \;\mbox{iff}\; \Delta(q_0,u) \in  Q_1$.

A \emph{Muller automaton} consists of a finite set of states $Q$,
a transition function $\Delta:Q \times A \to Q$, an initial state
$q_0\in Q$ and a family $\Ff$ of subsets of $Q$. A given infinite
word $u\in A^{\N}$ is accepted by a Muller automaton if the set of
states that are visited infinitely often by the path generated by
the given word is a member of $\Ff$. A language $L \subseteq
A^{\N}$ is \emph{$\omega$-regular}, if it is accepted by a Muller
automaton, i.e.,
$$u\in L \;\mbox{iff}\;
  \{q\in Q:\; \forall n,\exists m>n : \Delta(q_0,u_0\ldots,u_{m-1})
       =q\} \in \Ff.$$
Alternatively, $\omega$-regular languages can be defined by
nondeterministic B\"uchi finite automata. An infinite word is
accepted, if there is a trajectory passing infinitely often
through a given set of final states. Although B\"uchi automata are
simpler to define, Muller automata are deterministic, which is
sometimes an advantage. In this paper we make little use of
B\"uchi automata. Coming back to Figure \ref{fighalting1}, the
halting problem for a symbolic system asks whether there is a
finite word induced by the partition $U$,$V$, $W$ that is accepted
by the finite automaton. It is equivalent to ask whether there is
an infinite word induced by the partition that is accepted by the
automaton interpreted as a Muller automaton.

In general, given a clopen partition $\Aa=\{A_1,\ldots,A_N\}$ and
a finite automaton over $\Aa$, we would like to know whether there
is a non-empty intersection between the language associated to the
partition and the regular language accepted by the automaton. In
other words, the problem is to know whether there exists a point
of the symbolic system whose trajectory, when observed through the
partition, is accepted by the automaton. The same question can be
asked for a Muller automaton instead of a finite automaton.

The automaton may be interpreted as \emph{observing} the system.
This formalism includes all three properties described above,
including the halting problem.
 These are examples of \emph{model-checking}
problems. Model-checking aims at finding decision algorithms to
check whether the trajectories of a dynamical system satisfy a
given property. But systems considered in the literature of
model-checking are often non-deterministic and finite or
countable, whereas we deal with deterministic systems with a
possibly uncountable configuration space.

Note that Muller (or B\"uchi) automata are rather powerful to
express properties on infinite words. They are equivalent to
several logical formalisms, including the so-called $\mu$-calculus
and monadic second-order formulae. First-order formulae, including
(\ref{firstorderhalting}), (\ref{firstordervariant1}),
(\ref{firstordervariant2}), are equivalent to linear temporal
logic and strictly weaker than Muller automata. For precise
definitions of all these formalisms, see for instance
\cite{perrinpin,kaivola_thesis,modelchecking1999}.

\section{Decidable systems} \label{sectdecidability}

\begin{defi}  \label{defdecidab}
An  effective symbolic system is \emph{decidable} if there exists
an algorithm that decides the \emph{model-checking problem for
Muller automata}, i.e., that decides whether  the subshift induced
by a given clopen partition has a nonempty intersection with a
given $\omega$-regular language (described by a Muller automaton).
\end{defi}

Clearly, decidability is preserved by effective conjugacies and
the factor of a decidable system is decidable. The identity map on
any effective symbolic space is decidable. Indeed, for a partition
$A_1, A_2, \ldots, A_N$, the only words induced by the partition
are $A_1^\omega$, $A_2^\omega$, $\ldots$ and $A_N^\omega$. Given a
Muller automaton, it is enough to check whether one of these paths
starting from an initial state of the automaton passes infinitely
often through a final state. Alternatively it is a consequence of
the forthcoming Proposition \ref{equicont}. The map $x \mapsto 0x$
on $\{0,1\}^{\N}$ with a unique attracting fixed point
$0^{\omega}$ is decidable.  This follows from Proposition
\ref{attrminimal}. The full shift on any finite alphabet is a
decidable system by a corollary to Proposition \ref{shadow}.

If a system is not decidable, how undecidable can it be? We show
that the problem of model-checking is at most
$\Sigma_1^1$-complete, which is rather high. A \emph{$\Sigma_1^1$
set} is the set of integers $m$ satisfying a formula of the kind
$$\exists k, \, Q_1  n_1, \,
\ldots \, , Q_i n_i : R(k,m,n_1,\ldots,n_i),
$$
where $k$ runs over all total functions from $\N$ to $\N$,
$Q_1,\ldots, Q_i$ are quantifiers, $n_1, \ldots, n_i$ run over
$\N$, and $R$ is a recursive relation. By recursive we mean that
there is a Turing machine with $k$ as oracle and
$m,n_1,\ldots,n_i$ as data that decides in finite time whether
$R(k,m,n_1,\ldots,n_i)$ holds or not. A $\Sigma_1^1$ set is
\emph{$\Sigma_1^1$-complete} if  every $\Sigma_1^1$  is many-one
reducible to it. The class of $\Sigma_1^1$ problems belongs to the
so-called analytical hierarchy;  see \cite{rogers_book} for more
details.

\begin{prop} \label{prop_analytic_hier}
The problem of model-checking for Muller automata on an effective
symbolic system is at most $\Sigma_1^1$-complete.
\end{prop}

\begin{proof}
Let $f:X \rightarrow X$ be an effective symbolic system. First we
show that the model-checking problem is in $\Sigma_1^1$. Then we
construct a system simulating a universal Turing machine with
oracle for which the problem of model-checking is
$\Sigma_1^1$-complete. The proof, although rigorous, is not
completely formalized.

We can suppose that the space $X$ of the system is an effective
closed subset of the Cantor space $2^\N$. Let $x$ be a sequence
taking values in $\N$. Then the assertion `$x \in X$' is
equivalent to the recursive relation `$\forall t \in \N: x_0, x_1,
\ldots, x_t \in \{0,1\} \; \mathrm{and} \; [x_0 x_1 \ldots x_t]
\cap X \neq \emptyset$'. Let $m$ be a natural integer encoding a
B\"{u}chi automaton whose alphabet is a partition of $X$. Here
B\"{u}chi automata are of easier use than Muller automata. A
B\"{u}chi automaton is given by a finite set of states, an
alphabet and a transition relation, a set of initial states and a
set of final states. For any $x \in X$, call $R_f(x,m,t)$ the
relation `for the initial condition $x$, the B\"{u}chi automaton
$m$ observing the system can be in a final state at time $t$'. It
is a recursive relation; the configuration $x$ can be seen as a
function from $\N$ to $\N$. Then the problem of model-checking can
be expressed by the logical formula `$\exists x : x\in X \;
\textrm{and} \; \forall t, \exists t'\geq t: R_f(x,m,t')$', with
$m$ as free variable; hence model-checking for Muller automata is
in $\Sigma_1^1$.

The set of natural integers $n$ such that there exists a sequence
of integers $k: \N \rightarrow \N$ for which the universal Turing
machine with initial data $n$ and oracle $k$ does not halt is well
known to be $\Sigma_1^1$-complete; see \cite{rogers_book}. An
oracle universal Turing machine can be built in the following way.
We take a one-tape universal Turing machine in the usual sense, to
which we adjoin a tape that contains on its right part the oracle
encoded in form $10^{k(0)}10^{k(1)}10^{k(2)}1.\ldots$ The head has
access to both tapes. Not every possible content of the second
tape is a valid oracle; indeed the word $0^\omega$ cannot appear
on the tape. We can suppose without loss of generality that when
the head wants to query $k(i)$, it first checks that $k(i)$ is
properly encoded by scanning the tape in some state
$q_{\textrm{search}}$ until it discovers a $1$ and then jumps to
the state $q_{\textrm{found}}$. This two-tape Turing machine is an
effective dynamical system, similar to the one-tape Turing machine
discussed just above Section \ref{subsect shifts and subshifts}.
Call $Q$ the states of the head, $q_0$ the initial state and $q_h$
the halting state. It can be supposed that it is impossible to
leave $q_h$ once we reach it. We want to know whether there is an
initial configuration  of this system, composed of a state of $Q$
and the contents of both tapes, that is in the clopen set $\{q_0\}
\times [n] \times [1]$ (i.e., the head is in state $q_0$, the
initial data $n$ is encoded at the right of the head on the first
tape and a symbol $1$ is currently read by the head on the second
tape) and such that the head reaches infinitely often $Q \setminus
\{q_{\textrm{search}}, q_h \}$. For if an initial configuration is
such that the head does not reach infinitely often $Q \setminus
\{q_{\textrm{search}}, q_h \}$, then it either  reaches the
halting state or gets stuck in a query on an invalid oracle. This
property can be observed by a Muller automaton in a
straightforward manner. Putting all together, we have constructed
a reduction from a $\Sigma^1_1$-complete problem to a
model-checking problem of some fixed symbolic system; the latter
is therefore $\Sigma_1^1$-complete as well.
\end{proof}

\section{Universal systems} \label{sectuniversality}

We are now ready to state the main definition of computational
universality. We define a universal symbolic system as a special
kind of undecidable system, where Muller automata are replaced by
finite automata. The universality of Turing machines is a
particular example of this definition.

\begin{defi}  \label{defunivers}
An  effective dynamical system is \emph{universal} if the
\emph{model-checking problem for finite automata} of this system,
i.e., the problem whether the language induced by a given clopen
partition has a nonempty intersection with  a given regular
language, is recursively-enumerable complete.
\end{defi}

An \emph{r.e.-complete} problem, or \emph{$\Sigma_1$-complete}
problem, is a recursively enumerable problem, to which any
recursively enumerable problem is many-one reducible. Note that
the problem of model-checking for Muller automata (described in
Definition \ref{defunivers}) is always recursively enumerable,
because the language induced by a clopen partition is recursively
enumerable and the language accepted by a finite automaton is
recursive; the intersection can be recursively enumerated and if
it is nonempty then we can know it after a finite time.
Universality is obviously preserved by effective conjugacies, and
a system with a universal factor is also universal.

\begin{prop}
A universal system is not decidable.
\end{prop}

\begin{proof}
If the model-checking for Muller automata is decidable then so is
the model-checking problem for finite automata. Indeed, the latter
is reducible to the former in the following way. Given a
deterministic finite automaton, modify it in a such a way that the
final states are fixed points of the transition function, whatever
the input is; the resulting automaton is interpreted as a Muller
automaton, for the family of all sets whose unique elements is a
final state.
\end{proof}

Note that  a non-deterministic scheme of computation underlies the
definition of universality. The computation succeeds if and only
if at least one trajectory exhibits a given behavior. For example,
recall from Section \ref{sectautomata} that the halting problem
consists in determining, given the clopen sets $U$ and $V$,
whether there is a configuration  in $U$ that eventually reaches
$V$. We may think of $V$ as the halting set and of $U$ as an
initial configuration of which we know only the first digits. The
unspecified digits of the initial configuration may be seen as
encoding the non-deterministic choices occurring during the
computation.

\subsection{Examples} \label{subsectexamples}

\subsubsection*{Turing machines with blank symbol.}

A Turing machine with blank symbol that is universal in the sense
of Turing, is also universal according to Definition
\ref{defunivers}, because the halting problem `Can we go from a
clopen set $U$ to a clopen set $V$?' is r.e.-complete. Indeed the
halting problem restricted to clopen sets that are isolated points
is already r.e.-complete. Recall that isolated points are exactly
finite configurations. Incidentally, we have shown that what we
have called `halting problem' for a general symbolic system is
indeed a generalization of the usual halting problem for Turing
machines.

\subsubsection*{Turing machines without blank symbol.}
It is only slightly more complicated to build a universal Turing
machine without blank symbol. In such a Turing machine, there is
no obvious notion of `finite configuration'. The trick is
basically to encode the initial data in a self-delimiting way.
Take a Turing machine that is universal in the sense given by
Turing. Then add two new symbols $L$ and $R$ to the tape alphabet.
On an initial configuration, put an $L$ on the left end and an $R$
on the right end of the encoded data. When the head encounters an
$L$, it pushes it one cell to the left, leaving some more space
available for computation. It acts similarly for an $R$ symbol.
The working space is always delimited by an $L$ and an $R$; the
symbols situated outside this zone are considered as noise, and do
not influence the computation. For this modified universal Turing
machine, the (clopen-set-to-clopen-set) halting problem is again
undecidable.

\subsubsection*{Cellular automata.}
Let us take a universal Turing machine with a blank symbol. We
suppose that when the halting state is reached, then the head
comes back to the cell of index $0$. We can simulate it in a
classic way with a one-dimensional cellular automaton. The
alphabet of the automaton is $A \cup (A \times Q) \cup \{L, R,
Error\}$, where $A$ is the tape alphabet (including the blank
symbol) and $Q$ the set of states. Let us take a point in the
cylinder $[L,\text{initial data of the Turing machine}, R]$, and
observe its trajectory. The symbol $L$ moves to the left at the
speed of one cell per time step, leaving behind blank symbols. The
symbol $R$ moves to the right in a similar way. Meanwhile, the
space between $L$ and $R$ is used to simulate the Turing machine
and is composed of symbols from $A$ and exactly one symbol from $A
\times Q$, which denotes the position of the head. When $L$ or $R$
symbols meet each other, then a spreading $Error$ symbol is
produced, that erases everything.

This cellular automaton is again universal, because the
(clopen-set-to-clopen-set) halting problem is r.e.-complete.
Indeed, there is an orbit from the cylinder $[L,\text{initial data
of the Turing machine}, R]$ to the cylinder $[A \times
\{\text{halting state}]\}$ (both cylinders centered at cell of
index zero) if and only if the universal Turing machine halts on
the initial data.

\subsubsection*{Tag systems.}
Tag systems were introduced by Post in 1920. A \emph{tag system}
is a transformation rule acting on finite binary words. At every
step, a fixed number of bits is removed from the beginning of the
word and, depending on the values of these bits, a finite word is
appended at the end of the word. Minsky \cite{minsky1961} proved
that there is a so-called universal tag system, for which checking
whether a given word will eventually produce the empty word when
repeating the transformation is an r.e.-complete problem.

We can extend the rule of tag systems to infinite words, by just
removing from them a fixed number of bits. Thus we have a
dynamical system on the compact space $\{0,1\}^* \cup \{0,1\}^\N$
of finite and infinite words, in which finite words are clopen
sets. Again, if the tag system is universal for the word-to-word
definition, then it is universal for Definition \ref{defunivers}
with the halting problem on clopen sets of $\{0,1\}^* \cup
\{0,1\}^\N$.

\subsubsection*{Collatz functions.}

We can also apply our definition to functions on integers. Let $\N
\cup \{\infty\}$ be the topological space with the metric
$d(n,m)=|2^{-n}-2^{-m}|$. This is effectively homeomorphic to the
set $\{1^n0^\omega | n \in \N\} \cup \{1^\omega\}$. Then some
functions on integers may be extended to infinity. For instance,
the famous $3n+1$ function sends even $n$'s to $n/2$, odd $n$'s to
$3n+1$ and $\infty$ to $\infty$. Whether this map is decidable is
unsettled. But Conway \cite{conway_collatz} proved that similar
functions, called Collatz functions, can be universal.

\subsubsection*{Counter machines.}
A $k$-counter machine is  composed of $k$ counters, each
containing a non-negative integer, and a head that can test which
counters are at zero and can increment or decrement every counter
(with the convention $0-1=0$). Thus a counter machine is a map
$f:Q \times \N^k \rightarrow Q \times \N^k$, where $Q$ is the
finite set of states of the head. There exists such a machine $f$
for which given two configurations $x,y \in  Q \times \N^k$, the
problem to check whether the trajectory of $x$ reaches $y$ is
r.e.-complete; see Minsky \cite{minsky1961}.

The map $f$ is easily extended to the compact space $Q \times (N
\cup \{\infty\})^k$, with the convention $\infty \pm 1 = \infty$.
Here again, the points of $Q \times \N^k$ are clopen sets of $Q
\times (N \cup \{\infty\})^k$, hence $f$ is universal for the
halting problem.

\subsubsection*{More examples.}
In Section \ref{sectunivchaos} we give an example of a universal
system that is chaotic, and for which the halting problem is
decidable, but not the variant expressed by logical formula
(\ref{firstordervariant1}). In Section \ref{sectshadow} we build a
system which is neither decidable nor universal. In the setting of
point-to-point properties, it was proved  by Sutner
\cite{sutner_intermediateCA} that there exist cellular automata
with a halting problem of an intermediate degree between
decidability and r.e.-completeness. The same kind of examples for
Turing machines are known for long time (Friedberg-Muchnik
theorem, see for instance \cite{rogers_book}). However we have not
been able to build a system for which finite-automata properties
of trajectories are undecidable, but not r.e.-complete.

\section{Sufficient conditions for decidability}
\label{sectnecesscond}

The purpose of this section is to link computational capabilities
of a system to its dynamical properties. Most results proved in
this section are in fact sufficient conditions of decidability and
can thus be interpreted as necessary conditions for universality.
For instance, we prove that minimal systems are decidable, thus
universal systems are not minimal.


The following constructions and propositions are useful in several
proofs. Given an effective system $f:X \to X$, a clopen partition
$\Aa=\{A_1,\ldots,A_N\}$ of $X$ and the transition function
$\Delta:Q \times \Aa \to Q$ of a deterministic finite automaton,
we construct the \emph{observation system} $f_{\Delta}:X \times Q
\to X\times Q$ by
$$ f_{\Delta}(x,q) = (f(x),\Delta(q,A_i)),\;\mbox{where}\;
    x \in A_i $$
Clearly $f_{\Delta}$ is an effective system, and the projection
$\pi_X: X\times Q \to X$ is an effective factor map of
$f_{\Delta}$ to $f$.

\begin{defi}
We say that a dynamical system $f:X \to X$ has \emph{clopen
basins}, if for every clopen set $V \subseteq X$, its basin
$\Bb(V) = \bigcup_{n\geq 0} f^{-n}(V)$ is a clopen set.
\end{defi}

\begin{prop}
If $f:X \to X$ is an effective system with clopen basins, then the
operation $V \mapsto \Bb(V)$ is computable.
\end{prop}

\begin{proof}
If $V$ and $\Bb(V)$ are clopen sets, then by compactness there
exists $m>0$ such that
$$\Bb(V) = \bigcup_{n<m} f^{-n}(V) = \bigcup_{n<m+1} f^{-n}(V).$$
Given $V$ we can determine $m$ effectively so the operation
$\Bb(V)$ is effective too. Hence there exists a computable
function $k:\N \to \N$ such that $\Bb(P_n) = P_{k(n)}$, where
$P_n$ is the clopen set of index $N$.
\end{proof}

\begin{prop}  \label{basin}
If an effective system is such that for any transition function,
the resulting observation system has clopen basins, then the
system is decidable.
\end{prop}
\begin{proof}
For every clopen partition $\Aa$, for every finite set $Q$ and for
every transition function $\Delta: Q \times \Aa \to Q$, the system
$f_{\Delta}:X \times Q \to X \times Q$ has clopen basins.

Assume now that $V\subseteq X\times Q$ is clopen, so that $\Bb(V)$
is clopen and the index of $\Bb(V)$ can be computed from the index
of $V$. Moreover $I(V)$, defined as $\Bb(\Bb(V)^c)^c$, where $^c$
denotes the complement, is a clopen set too and its index can be
again computed from that of $V$. A point $(x,q)$ belongs to $I(V)$
iff the trajectory of $(x,q)$ passes through $V$ infinitely often.
Given $q_0,q_1 \in Q$, then $(X\times \{q_0\})\cap I(X\times
\{q_1\})$ is again a computable clopen set, so the set $\{x\in
X:\; \forall n,\exists m>n, f_{\Delta}^m(x,q_0)=q_1\}$ is
computable as well. It follows that for a family $\Ff$ of subsets
of $Q$, the set
$$ \{x\in X:\; \{q\in Q: \; \forall n,\exists m>n, f_{\Delta}^m(x,q_0)=q\}
       \in \Ff\}       $$
is computable too. In particular, whether this set is empty can be
decided algorithmically. Hence the model-checking problem for
Muller automata is decidable.
\end{proof}

\subsection{Minimality}

We first prove that an undecidable system must have a `thin'
subsystem.

\begin{prop} \label{prop_nonempt_int_decid}
A symbolic system such that all nonempty subsystems have a
nonempty interior is decidable.
\end{prop}
\begin{proof}
Consider an effective symbolic dynamical system $f:X \rightarrow
X$ whose subsystems have a nonempty interior, a partition
$\Aa=\{A_1, A_2, \ldots, A_N\}$ of $X$ and consider a Muller
automaton whose set of states is $Q$, alphabet is $\{A_1, A_2,
\ldots, A_N\}$ and transition function is $\Delta: Q \times \Aa
\to Q$.

We check that the observation system has no subsystem with an
empty interior (except the empty set). Indeed let $Y \subseteq X
\times Q$ a closed subset such that $f_\Delta(Y) \subseteq Y$. For
every $q \in Q$, call $Y_q \subseteq X$ the projection on $X$ of
$Y \cap X \times \{q\}$. Then the projection of $Y$  is the union
of all $Y_q$ and is a subsystem of $X$. By hypothesis, it has a
nonempty interior; thus one of the $Y_q$ has a nonempty interior
(this follows from Baire's theorem). This implies that one of the
sets $Y \cap X \times \{q\}$ has a nonempty interior. Hence $Y$
itself has a nonempty interior in $X \times Q$.

We want to show that the observation system $f_\Delta$ has clopen
basins.

Given a clopen set $V \subseteq X \times Q$, we show that the
border $\partial\Bb(V)$ is $f_\Delta$-invariant. Indeed
$f_\Delta(\Bb(V)\setminus V)\subseteq \Bb(V)$ and, since $V$ is
clopen, $\partial\Bb(V) =
\partial(\Bb(V)\setminus V)$. If $y \in
\partial\Bb(V)$ and $U$ is a neighborhood of $f_\Delta(y)$, then
$f_\Delta^{-1}(U)$ is a neighborhood of $y$ that contains a point
$z \in \Bb(V)\setminus V$. Thus $f_\Delta(z) \in U \cap \Bb(V)$.
Since $U$ is arbitrary,  this proves that $f_\Delta(y) \in
\partial \Bb(V)$. But $f_\Delta(y)\not\in \Bb(V)$, because $y\not\in
\Bb(V)\setminus V$. This proves the invariance of
$\partial\Bb(V)$.

Since $\partial\Bb(V)$ is a subsystem with empty interior, it must
be empty, so $\Bb(V)$ is clopen. So the observation system has
clopen basins. By Proposition \ref{basin}, this proves the system
$f$ to be decidable.
\end{proof}

A \emph{minimal} dynamical system is a system with no subsystem
(except the empty set and itself). In minimal system, all orbits
are dense and the basin of any clopen set is the full set. A
minimal system trivially satisfies the hypothesis of the preceding
proposition, and so we have the following.

\begin{prop} \label{minimal}
A minimal symbolic system is decidable.
\end{prop}

This is in a way not surprising since in some way all trajectories
of a minimal system have the same behavior.

Any dynamical system has a minimal subsystem, thanks to Zorn's
lemma and compactness. In particular, any point comes arbitrarily
close to a minimal system, since the closed orbit of the point is
itself a dynamical system. Suppose that the symbolic system is not
minimal but consists of one  minimal subsystem attracting the
whole space of configurations. In other words, the limit set is
minimal. The limit set of a dynamical system $f:X \rightarrow X$
is the set $\bigcap_{n \geq 0} f^{n}(X)$. Then such a system is
again decidable. This results from the more general following
proposition.

\begin{prop} \label{attrminimal} \label{severalattrminimal}
A symbolic system whose limit set is the union of finitely many
minimal systems is decidable.
\end{prop}
\begin{proof}
Given a symbolic system $f: X \rightarrow X$ and a Muller
automaton whose set of states is $Q$, we build the observation
system $f_\Delta: X \times Q \rightarrow  X \times Q$.

First we prove that the observation system $f_\Delta$ contains
finitely many minimal sets. Let $X_1,\ldots,X_k$ be the minimal
subsystems of $f: X \to X$. For every $i=1,\ldots,k$ choose an
arbitrary point $x_i \in X_i$. A minimal subsystem of $f_\Delta$,
when projected on $X$, is exactly a minimal subsystem of $f$, as
easily seen. Thus any minimal subsystem of $f_\Delta$ must contain
at least one point of the form $(x_i,q)$, for some $q \in Q$.
Since any two different minimal subsystems are disjoint, this
means that there are at most $k|Q|$ minimal subsystems in
$f_\Delta$.

Then we show that the limit set of $f_\Delta$ is exactly the union
of all minimal subsystems.

It is clear that the minimal subsystems are in the limit set of
$f_\Delta$. Now we prove that each minimal subsystem $Z$ of
$f_\Delta$ has a nonempty interior in the limit set of $f_\Delta$
(for the relative topology). The projection of the limit set of
$f_\Delta$ on $X$ is the limit set of $f$. The projection of $Z$
on $X$ is a minimal subsystem of $f$, which has a nonempty
interior in the limit set of $f$, and the projection of $Z$ is
$\bigcup_{q \in Q} Z_q$, where $Z=\bigcup_q Z_q \times \{q\}$.
From Baire's theorem, one of these $Z_q$ has a nonempty interior
in the limit set of $f$, and $Z$ itself has a nonempty interior in
the limit set of $f_\Delta$.

Let $Y_i$ be a set included in $Z_i$ that is open in the limit set
of $f_\Delta$, where  $Z_1,\ldots,Z_m$ are the minimal subsystems
of $f_\Delta$. All sets $\bigcup_{n \in \N} (f_\Delta^{-n}(Y_i))$
are disjoint sets, are open in the limit set and cover the limit
set, since the closed orbit of every point in the limit set of
$f_\Delta$ must include a minimal subsystem. From compactness, all
points of the limit set of $f_\Delta$ fall in a minimal subsystem
in  bounded time. We conclude that the union of all minimal
subsystems is the exactly the limit set of $f_\Delta$.

So the limit set of the observation system $f_\Delta$ is a finite
union of minimal subsystems. We get from the lemma below that
$f_\Delta$ has clopen basins. From Proposition \ref{basin} we
deduce that $f$ is decidable.
\end{proof}

For instance, the system $f: \{0,1\}^\N \to \{0,1\}^\N: x \mapsto
0x$ is decidable. The following lemma finishes the proof.

\begin{lemm}
A symbolic system whose limit set is the finite union of minimal
systems has clopen basins.
\end{lemm}
\begin{proof}
Suppose that the limit set is $Y_1 \cup\cdots\cup Y_k$, where
$Y_i$ are minimal subsystems, so that $Y_i\cap Y_j=\emptyset$ for
$i\neq j$. Let $V \subseteq X$ be a clopen set. If $V\cap Y_i
=\emptyset$, then $\Bb(V) \cap Y_i = \emptyset$. If $V \cap Y_i
\neq \emptyset$, then  for some $m>0$, $Y_i \subseteq V_m =
\bigcup_{n<m} f^{-n}(V)$. Thus there exists $m>0$ such that for
all $i$ either $Y_i \subseteq V_m$ or $Y_i\cap V_m=\emptyset$.
Then $W_m = f^{-m}(V)\setminus V_m$ is a clopen set disjoint from
the limit set. From compactness there exists $k>0$ such that
$f^{-k}(W_m) =\emptyset$, so $\Bb(W_m)$ is a clopen set. It
follows that $\Bb(V) = V_m \cup \Bb(W_m)$ is a clopen set too.
\end{proof}

The above proposition is in fact more general than Proposition
\ref{prop_nonempt_int_decid}:

\begin{prop}
A symbolic system such that all nonempty subsystems have a
nonempty interior has a limit set composed of finitely many
minimal subsystems.
\end{prop}
\begin{proof}
Let $f$ be a system such  that all nonempty subsystems have a
nonempty interior. In the interior of every minimal subsystem
choose a clopen set $U_i$. The basin of the open set $\bigcup_i
U_i$ is the full space, because every point of the system must
come arbitrarily close to some minimal subsystem, thus must fall
in some $U_i$. By compactness, there is a finite set of $i$s and a
natural integer $m$ such that $\bigcup_{i \in I} \bigcup_{n < m}
f^{-n}(U_i)$ is the full space. So there are finitely many minimal
subsystems, and every point falls in a finite time into a minimal
subsystems. The union of the minimal subsystems is therefore the
limit set.
\end{proof}

A stronger statement than Proposition \ref{attrminimal} is
suggested by the intuition that an undecidable system (and
especially a universal system) is likely to be able to `simulate'
many other systems.

\begin{conj} \label{conjinfinite}
A universal symbolic system has infinitely many minimal subsystems.
\end{conj}

\subsection{Equicontinuity}

A system $f: X \rightarrow X$ is \emph{equicontinuous} if for
every $\epsilon > 0$ there exists a $\delta > 0$ such that $d(x,y)
< \delta$ implies $d(f^t(x),f^t(y)) < \epsilon$, for any points
$x,y$ and $t \in \N$. Note that equicontinuity in symbolic systems
is a topological property not just a metric one. Instead of `For
every $\epsilon >0$, there is a $\delta$ \ldots' we could say `For
every clopen partition, there is a finer clopen partition such
that if two points are in the same subset of the finer partition,
then they generate the same infinite word in the coarser
partition.'

\begin{prop}  \label{equicont}
An equicontinuous symbolic system is decidable.
\end{prop}

\begin{proof}
First we prove that an equicontinuous system has clopen basins.
Let $V$ be a clopen set. Then from equicontinuity there exists a
$\delta$ such that any two points distant of less than $\delta$
either both eventually reach $V$ or both never reach $V$. Hence
$\Bb(V)$ is the union of balls of radius $\delta$, and is a clopen
set.

Then we show that the observation system of an equicontinuous
symbolic system $f: X \to X$ is equicontinuous. The space $X
\times Q$ (for a finite set $Q$) can be endowed the metric
$d((x,q),(x',q'))=d(x,x')$ if $q = q'$ and $1$ otherwise. For a
partition $\Aa=\{A_1,\ldots,A_n\}$ of $X$, let $\zeta >0$ be such
that every $A_i$ is a union of open balls of radius $\zeta$. Then
for any $\epsilon$, choose a $\delta$ such that any two points of
$X$ distant of less than $\delta$ have orbits distant of less than
$\min(\epsilon,\zeta)$. Then two points of the observation system
distant of less than $\delta<1$ have the same second component,
will always be in the same partition subset and will always have
the same second component. Hence their distance will always be
less than $\epsilon$. Consequently, the observation system is
equicontinuous.

It follows by Proposition \ref{basin} that the system is
decidable.
\end{proof}

We say that a point $x$ of a dynamical system $f$ is
\emph{sensitive} if there is an $\epsilon> 0$ such that for every
$\delta >0$ there is a point $y$ with $d(x,y) < \delta$ and a
non-negative time $t$ such that $d(f^t(x), f^t(y)) > \epsilon$. It
is easy to show with compactness that an equicontinuous dynamical
system is exactly a system with no sensitive point. Hence,
Proposition \ref{equicont} implies that an undecidable symbolic
system must have a sensitive point. Equicontinuity in the case of
cellular automata has been given a combinatorial characterization
in \cite{kurka_equicont_cellaut}, where it is also proved that
equicontinuous cellular automata are eventually periodic, thus
confirming in this particular case  that equicontinuity is
incompatible with computational universality.

\subsection{Regular Systems}

A subshift is called sofic, if its language is regular. A symbolic
system is called \emph{regular}, if all its induced subshifts are
sofic; see \cite{kurka_simplicity,kurka_universality}. Can a
regular system be universal? We first consider a closely related
question. We say that an effective system is \emph{effectively
regular} if it is regular and there is an algorithm that builds
from a given clopen partition the finite automaton recognizing the
regular language induced by the partition.

\begin{prop} \label{prop eff regular}
An effectively regular system is decidable.
\end{prop}

\begin{proof}
The intersection of two $\omega$-regular languages is well known
to be an $\omega$-regular language, and a Muller automaton
accepting the intersection can be computed; see \cite{perrinpin}
for instance. Moreover, whether the language accepted by a given
Muller automaton is empty is a decidable problem too. And a sofic
subshift is an $\omega$-regular language: the finite automaton
accepting the language, interpreted as a Büchi automaton with the
same set of final states, accepts the sofic subshift.

Suppose that we are given an effectively regular system, a clopen
partition $\Aa$ of the space and a Muller automaton over the
alphabet $\Aa$. Then we construct another Muller automaton that
accepts exactly the subshift induced by $\Aa$ and verify whether
the languages accepted by these two Muller automata has a nonempty
intersection. Hence the system is decidable.
\end{proof}

If the system is regular but not effectively regular, then the
argument of the proof fails.

\begin{prop}  \label{soficuniversal}
There exists a symbolic system that is regular and universal.
\end{prop}

\begin{proof}
Let $X_n$ be the subshift  of $\{0,1\}^{\N}$ whose forbidden words
are words of the form  $10^t1$, where $t$ is less than the
(possibly infinite) halting time of the universal Turing machine
launched on data $n$. If the Turing machine does not halt, then
$X_n$ is the sofic subshift $\{0^*10^\omega, 0^\omega\}$. If the
Turing machine halts in $k$ steps, then $X_n$ is the subshift of
finite type with forbidden words $11$, $101$, $1001$, \dots,
$10^{k-1}1$. So all subshifts are sofic, but we cannot effectively
build the automaton recognizing the language, for it would allow
to solve the halting problem.

Now consider the product of all $X_n$. This product is again an
effective symbolic system $X$, and all its induced subshifts are
sofic, due to the fact that the finite product of sofic subshifts
is a sofic subshift and the induced subshift of a sofic subshift
is again sofic; see \cite{kurka_book}. Thus the system is regular,
but not effectively regular. Finally, it is r.e.-complete to check
whether there is a trajectory starting from $\pi_n^{-1}([1])$
which eventually reaches $\pi_n^{-1}([01])$. Here $\pi_n:X \to
X_n$ is the projection.
\end{proof}

\subsection{Shadowing property} \label{sectshadow}

\begin{defi} \label{defshadow}
Let  $f: X \rightarrow X$ be a symbolic dynamical system. A
\emph{$\delta$-pseudo-orbit} is a (finite or infinite) sequence of
points $(x_n)_{n \geq 0}$ such that $d(f(x_n), x_{n+1}) < \delta$
for all $n$. A point $x$ \emph{$\epsilon$-shadows} a (finite or
infinite) sequence $(x_n)_{n \geq 0}$ if $d(f^n(x), x_n) <
\epsilon$ for all $n$.
A dynamical system is said to have the \emph{shadowing property}
if for every $\epsilon > 0$ there is a $\delta >0$ such that any
$\delta$-pseudo-orbit is $\epsilon$-shadowed by some point. If
moreover such a rational $\delta$ can be effectively computed from
a rational $\epsilon$ then we say that the system has the
\emph{effective shadowing property}.
\end{defi}

For example, the one-sided and two-sided shifts have the shadowing
property for $\delta=\epsilon$. By a theorem of Walters, a
subshift of finite type has the shadowing property, with a linear
relation between $\epsilon$ and $\delta$ (see \cite{kurka_book}
for a proof), thus has the effective shadowing property. Clearly,
the effective shadowing property is invariant under effective
conjugacies. We can give the following interpretation to the
effective shadowing property. Suppose that we want to compute
numerically the trajectory of $x$ such that at every step
numerical errors are bounded by $\delta$. The resulting sequence
of points is a $\delta$-pseudo-orbit, and the shadowing property
ensures that this pseudo-orbit is $\epsilon$-close to an actual
trajectory of the system, ensuring that the result of the
numerical computation is not meaningless.

\begin{prop} \label{lemmshadregular}
A symbolic system (effective or not) with the shadowing property
is regular. An effective symbolic
system with the effective shadowing property is effectively regular.
\end{prop}

\begin{proof}
The proof generalizes Proposition 5.69 of \cite{kurka_book} about
cellular automata. Consider a symbolic system $f: X \rightarrow X$
with the shadowing property and a clopen partition
$\Aa=\{A_1,\ldots,A_N\}$. There exists an $\epsilon$ such that all
clopen sets of the partition are finite unions of balls of radius
$\epsilon$. By the  shadowing property, there exists $\delta$ such
that every $\delta$-pseudo-orbit is $\epsilon$-shadowed. We may
suppose without loss of generality that $\delta \leq \epsilon$.
Let $\Bb=\{B_1,\ldots,B_M\}$ the clopen partition where each $B_i$
is a ball of radius $\delta$. Then the set of of all infinite
words induced by all $\delta$-pseudo-orbits through $\Bb$ is a
subshift of finite type: the word $B_i B_j$ is forbidden iff $B_i
\cap f^{-1}(B_j) = \emptyset$, i.e., we cannot go from $B_i$ to
$B_j$ in one step. But the partition $\Aa$ is coarser than $\Bb$,
so the subshift induced by $\Aa$ is a factor of a subshift of
finite type, hence sofic. If the system has the effective
shadowing  property, then we can effectively find $\delta$,
effectively describe the subshift of finite type and effectively
build the sofic subshift.
\end{proof}

\begin{prop} \label{shadow}
A symbolic system that has the effective shadowing property is
decidable.
\end{prop}

\begin{proof}
By Propositions \ref{lemmshadregular} and \ref{prop eff regular}.
\end{proof}
In  particular, the shift and any subshift of finite type is
decidable. Proposition \ref{shadow} is stronger than Proposition
\ref{equicont}, as we now show.

\begin{prop}
An equicontinuous effective symbolic system has the effective
shadowing property.
\end{prop}

\begin{proof}
Let $f: X \rightarrow X$ be an equicontinuous system. Then for
every $\epsilon>0$, there is a $\delta$ such that any two points
distant of less than $\delta$ have $\epsilon$-close trajectories.
We show that any $\delta$-pseudo-orbit is $\epsilon$-shadowed by
some point.

Let $x_0, x_1, x_2, \ldots$ be a $\delta$-pseudo-orbit. We show by
induction on $m$ that $d(f^{n}(x_m), f^{n+m}(x_{0})) < \epsilon$
for every $m$ and $n$. The case $m=0$ is obvious. If it is true
for $m$ then $d(f^{n+1}(x_m), f^{n+m+1}(x_{0})) <
\epsilon$. But $d(x_{m+1}, f(x_{m})) < \delta$ implies
$d(f^{n}(x_{m+1}), f^{n+1}(x_{m})) < \epsilon$. From the
ultrametric inequality we have $d(f^{n}(x_{m+1}),
f^{n+m+1}(x_{0})) < \epsilon$.

It is now enough to prove that a suitable $\delta$ is computable
from $\epsilon$, i.e. an equicontinuous symbolic system is always
`effectively' equicontinous. Take the partition $\Bb_0$ of all
balls of radius $\epsilon$. For $n=0,1,2, \ldots$, let $\Bb_{n+1}$
be the coarsest partition finer than $\Bb_{n}$ and
$f^{-1}(\Bb_{n})$. From equicontinuity, this sequence of finer and
finer partitions must stabilize to some
$\Bb_{n}=\Bb_{n+1}=\Bb_{n+2}=\cdots$. To check that we have
reached this point it is enough to check that $\Bb_{n}=\Bb_{n+1}$.
We choose $\delta$ so that the clopen sets of $\Bb_{n}$ can be
expressed as balls of radius $\delta$.\end{proof}

We also have the following result.

\begin{prop} \label{prop_shad_not_universal}
A symbolic system that has the shadowing property is not
universal.
\end{prop}

\begin{proof}
Let $f: X \rightarrow X$ be a symbolic system with the shadowing
property. Given a deterministic finite automaton observing the
system through a given clopen partition, the problem is to check
whether there exists a finite word induced by the clopen partition
that is accepted by the automaton. As we have noticed after
stating Definition \ref{defunivers}, this problem is recursively
enumerable. We show that it is also co-recursively enumerable.
This will prove that the problem is  decidable and that $f$ is not
universal.

Let $\Aa= \{A_1,\ldots,A_N\}$ be a clopen partition and $\Delta: Q
\times \Aa \rightarrow Q$ the transition function of a
deterministic finite automaton. We must essentially prove that the
halting problem is decidable for the observation system
$f_{\Delta}:X\times Q \to X\times Q$.

But $f_{\Delta}$ is an effective symbolic system with the
shadowing property, as we now show. We can suppose that the
distance between $(x,q)$ and $(x',q')$ is $1$ if $q \neq q'$ and
$d(x,x')$ otherwise. For an $\epsilon >0$, choose an $\epsilon'
\leq \epsilon$ such that any $A_i$ can be written as a union of
balls of radius $\epsilon'$. Then the shadowing property for $f$
yields a corresponding $\delta'$. Choose a $\delta \leq \delta'$
such that $\delta$ is strictly smaller than the distance between
any two sets $X \times \{q\}$ and $X \times \{q'\}$. Then it is
easy to see that any $\delta$-pseudo-orbit of $f_{\delta}$ is
$\epsilon$-shadowed by some point of $X \times Q$: such a
pseudo-orbit is projected onto a $\delta$-pseudo-orbit of $f$,
which is $\epsilon$-shadowed by some point, and this point can be
lifted to a point that $\epsilon$-shadows the pseudo-orbit of
$f_\Delta$.

Take two clopen sets $U,V \subseteq X\times Q$. There exists an
orbit from $U$ to $V$ iff for every $\delta>0$ there exists a
$\delta$-pseudo-orbit from $U$ to $V$ (see Proposition 2.15 of
\cite{kurka_book}). If there is no orbit starting in $U$ that
reaches $V$, then there exists a $\delta$ such that no
$\delta$-pseudo-orbit goes from $U$ to $V$, and we can
algorithmically check it by the following method. For a fixed
$\delta$, define $V'$ as the union of balls of radius $\delta$
whose center is in $f_\Delta^{-1}(V)$. Then compute $V''$, $V'''$,
and so on. As there are  only finitely many balls of radius
$\delta$, $V^{(t)}=V^{(t+1)}$ for some $t$. Then check whether
$V^{(t)} \cap U$ is empty; it is the case if and only if there is
no $\delta$-pseudo-orbit from $U$ to $V$. Start again with smaller
and smaller $\delta$.

Thus the halting problem for $f_\Delta$ is decidable. In
particular if $U= X \times \{q_0\}$ (where $q_0$ is the initial
state of the automaton) and $V= X \times F$ (where $F \subseteq Q$
is the set of final states of the automaton), then we can
algorithmically check whether there exists a  point of $X$ which
induces through the clopen partition a word that is accepted by
the automaton.
\end{proof}

The following proposition shows that the  effective shadowing property
is stronger than the shadowing property.

\begin{prop}  \label{propshadow_but_not_eff}
There exists an undecidable effective symbolic system that has the
shadowing property, but not the effective shadowing property.
\end{prop}

\begin{proof} Let $X_n$ be the subshift with forbidden words $0^t$, where the
universal Turing machine  stops on data $n$ in at most $t$ steps.
If the Turing machine does not halt on $n$, then $X_n$ is the full
shift; if it stops in $k$ steps, then the forbidden word is $0^k$.
All these subshifts are effective, but we cannot compute their set
of forbidden words.

The product $X$ of all $X_n$ is an effective system. Whether there
is a point that remains for ever in $\pi_n^{-1}[0]$ is
co-r.e.-complete (where $\pi_n: X \rightarrow X_n$ is the
projection). This property has been shown in Figure
\ref{fighalting3} to be expressible in terms of Muller automata.
Hence the system is undecidable.

Recall that a subshift of finite type has the shadowing property.
We show that the countable product of subshifts that have the
shadowing property also has the shadowing property. A ball of
radius $\epsilon$ in the product system may be expressed as the
finite union of products of balls of radius $\epsilon'$ in a
finite number of constituent subshifts. We choose the smallest of
the corresponding $\delta'$ given by shadowing property in the
subshifts. The product of balls of radius $\delta'$ may be
expressed as union of balls of radius $\delta$; this is the
$\delta$ corresponding to $\epsilon$.

Hence the system $X$ has the shadowing property but not the
effective shadowing property, since it is undecidable.
\end{proof}

As the shadowing property implies non-universality, it also proves
that universality is stronger than undecidability.

\begin{coro} There exists a symbolic system that is neither
decidable nor universal.
\end{coro}

Note also that Turing machines that satisfy the shadowing property
have been given a combinatorial characterization in
\cite{kurka_topo_TM}; in particular, the proof shows that the link
between $\epsilon$ and $\delta$ (see Definition \ref{defshadow}) is
linear. Hence the effective shadowing property is not stronger than
the shadowing property in the case of Turing machines.

\section{A universal chaotic system}
\label{sectunivchaos}

According to Devaney \cite{devaney}, a system is \emph{chaotic} if
it is infinite, topologically transitive and has a dense set of
periodic points. By \emph{topologically transitive} we mean that
for any two open sets $U$ and $V$, there is a point of $U$ that
eventually reaches $V$. One can prove that every point of a
chaotic system is sensitive \cite{banks_sensitiv_in_chaos}. For
instance, the full shift is chaotic and sensitive in every point.

It is not difficult to construct a universal subshift. Indeed, in
$\{0,1\}^{\N}$ consider all forbidden words of the form
$01^n00^t1$, where the universal Turing machine launched on  data
$n$ does not halt in less than $t$ steps. Then the subshift of all
configurations avoiding this set of words is effective and
universal: the halting problem is r.e.-complete. Modifying this
construction, we get the following result:

\begin{prop}
There exists an effective system on the Cantor space that is
chaotic and universal.
\end{prop}

\begin{proof}
Consider a subshift $X\subset \{0,1,\S\}^{\N}$ whose forbidden
words are all $01^n00^t1$, where the universal Turing machine
launched on  data $n$ does not halt in less than $t$ steps. Denote
by $L \subset \{0,1\}^*$ the language of binary words with no
forbidden subword. Then the language of $X$ consists of words
$w_1\S w_2\S \ldots \S w_n$, where $w_i\in L$. We show that $X$ is
a universal chaotic system.

First note that $X$ is a perfect subshift, so it is effectively
conjugated to a system on the Cantor space. Then $X$ has dense
periodic points: if $w\in L$, then $(w\S)^\omega$ is in $X$.
Finally $X$ is topologically transitive: for any two finite words
$v,w$ of the language we can go from $[v]$ to $[w]$ with the point
$v\S w.\ldots$ Thus $X$ is chaotic.

Moreover, given $n$ it is undecidable whether there is a point of
$[01^n0]$ that eventually reaches $[001]$ without passing through
$[\S]$. This property can be expressed by the finite automaton
constructed in Figure \ref{fighalting2}. Thus $X$ is universal.
\end{proof}

Note that the system built in the proof is a one-sided subshift,
hence it is expansive: there is an $\epsilon$ such that any two
points are eventually separated by at least $\epsilon$. Note also
that the halting property is decidable for

The central idea of the `edge of chaos' is that a system that has
a complex behavior should be neither too simple nor chaotic. There
are several ways to understand that.
Here we interpret `complex system' by `universal symbolic system'.
Then `too simple' could refer to the situation treated in
Proposition \ref{attrminimal}: one or several attracting
minimal subsystems. This includes of course the case of a globally
attracting fixed point.
If we take `chaotic' as meaning `Devaney-chaotic', then
computational universality need not be on the `edge of chaos',
since we have just constructed a chaotic system that is universal.

However, many examples of chaotic  systems (whatever the exact
meaning given to `chaotic', and for symbolic systems as well as
for analog ones), have the shadowing property. For instance the
shift and Smale's horseshoe (present in some physical systems), as
well as  hyperbolic systems, satisfy the shadowing property.

Thus we suggest that the term `edge of shadowing property' would
be more appropriate (at least for symbolic systems), although not
as thrilling.

Note nevertheless that the `edge of chaos' has been much studied
in cellular automata, and we don't know whether an example of a
chaotic universal cellular automaton exists.

\section{Discussion of universality}
\label{sectdiscussdef}

Turing \cite{turing} justified the form of his machine along the
following lines. A human operator applying an algorithmic
procedure can be supposed to be at every step of time in a unique
mental state. He can be supposed to have finitely many possible
mental states, and to have at his disposal a pencil and as much
paper as needed, on which he may write out letters or digits. In a
finite time he may read or write only finitely many symbols on the
paper. Paper is modelled by the tape and the human by a kind of
finite automaton that is able to read, write or shift the tape.

Now suppose that the human operator has no paper or pencil, but
can observe a (physical realization of) a symbolic dynamical
system, without being able to control it. The system can serve as
a `universal computer' if with its help, the human operator is
able to solve all problems he could also solve with paper and
pencil. As the human operator has finitely many possible mental
states, at every step he can distinguish only finitely many
configurations of the system. If we group together all points that
are undistinguishable between them, we obtain a partition of the
system state space. We suppose that this partition is clopen,
because clopen partitions express in a natural way that finitely
many symbols are observed from the system at every step of time,
analogously to Turing's assumption.

A symbolic system can be used as a `computer', if it is
computationally universal. When we observe the system using a
given finite automaton acting on a clopen partition, then deciding
whether the finite automaton can reach a final state from an
initial state is at least as difficult as deciding the halting
problem for a universal Turing machine. This means that when we
look for the answer to a recursively enumerable problem, then we
can obtain the answer by observing the system, provided we are
`lucky' and wait long enough.

Our definition of universality perhaps differs in several ways
from what we could  expect at first glance from a generalization
of Turing machine universality. We give now various arguments to
support the present definition against seemingly more obvious
attempts. In particular, we justify the use of \emph{set-to-set}
properties, observed by \emph{finite automata}, on systems defined
by a \emph{computable} map.

\subsection{Set-to-set properties} \label{pointtopoint}

Davis \cite{davis_universality} proposed the following definition:
a Turing machine is universal if the relation `$x_n$ is in the
orbit of $x_m$' is r.e.-complete, where $x_m$ and $x_n$ are
arbitrary finite configurations. This definition has the advantage
to bypass the need for a description of a way to encode the input
and decode the output of a computation.
Many definitions of universality for particular systems (cellular
automata, for instance) propose to observe point-to-point
properties.

Hemmerling \cite{hemmerling} proposes a definition for an
effective metric space; the basic idea is to endow a metric space
with a countable dense set of points. Examples include the reals
with rational points, the Cantor space with ultimately constant
configurations, the Cantor space with ultimately periodic
configurations.
This seems to provide a suitable framework to generalize Davis'
definition. Let us say that a metric space endowed with a dense
set of points $(x_n)_{n\in \N}$ is universal if the property
`$x_n$ is in the trajectory of $x_m$' is r.e.-complete.

However, as remarked in \cite{durand_roka_gameoflife}, this leads
to conclude that the shift is universal; a consequence that is
counter-intuitive. Indeed, consider the set of all configurations
with primitive recursive digits. This set is countable and dense.
Then we take as an initial configuration the sequence of pairs
(state of the head, currently read symbol) of a universal Turing
machine during a computation. And we only have to shift it to know
whether the halting state will ever appear. It sounds unreasonable
to classify the shift among universal systems, because it does not
compute anything but just reads the memory.

The definition presented in Section \ref{sectuniversality}
overcomes this problem in a simple manner: the user needs only to
specify a finite number of bits as an initial condition. Instead
of initial \emph{configurations} we should rather talk about
initial \emph{sets}, which may be seen as `fuzzy points', points
defined with finite accuracy.
This solution is also more satisfactory from the point of view of
physical realizability. Indeed, we expect the set of
configurations of a physical system to be uncountable in general,
and specifying an initial point for the computation means \emph{a
priori} that we must give an infinite amount of information.
Preparing a physical system to be in a very particular
configuration is likely to be impossible, because of the noise or
finite precision inherent to every measure.

\subsection{Finite automata}

What kind of property are we going to test on clopen sets (or,
equivalently, on induced subshifts)? Here again, we must avoid
trivialities. Suppose that we look at identity on the Cantor
space. We now choose to observe  the following property: a clopen
set satisfies the property if and only if its index (i.e., the
integer describing the clopen set) satisfies some r.e.-complete
property on $\N$. Then we find that the identity is computationally
universal, which is a result not to be desired. The complexity of
computation is artificially hidden in the decoding.

On the other hand, we see no reason to restrict ourselves to the
sole halting property: `there is a trajectory from this clopen set
to that clopen set'. Any observable property could \emph{a priori}
be used as a basis for computation. For instance, the chaotic
system built in Section \ref{sectunivchaos} is universal but the
halting property is decidable. So we must precisely define a class
of observable properties of clopen sets, not too large and not too
restricted. Finite automata used to express properties of finite
words have been extensively studied in the literature. They also
agree with Turing's idea of modelling a human operator as having
finitely many possible mental states. We do not use the powerful
setting of Muller automata to define universality, because it may
need an infinite time to check that a trajectory has the required
property, which goes against the idea that a successful
computation should end in a finite time. Whether a given observer
Muller automaton accepts at least one trajectory of the system is
actually a more general question, which is dealt with in our
definition of `decidable system'.

\subsection{Effectiveness}

Finally, the following example shows that  it is useful to add an
effectiveness structure on dynamical systems. Fix an r.e.-complete
set $H \subset \N$ of integers and consider the symbolic system
$f: \{0,1\}^\N   \rightarrow  \{0,1\}^\N$ such that $f(1^{\omega})
= 1^{\omega}$ and $f(1^n0x_0x_1x_2\ldots) = 1^m0x_0x_1x_2\ldots$,
where $m$ depends on $n$ in the following way. If $n\in H$, then
$m$ is the largest integer strictly smaller than $n$ such that $m
\in H$ or $0$ if no such number exists. If $n \not\in H$, then
$m=n$. Suppose now that $13 \in H$. Then the relation `the clopen
set $[1^n0]$ will eventually reach  $[1^{13}0]$' is r.e.-complete,
because $H$ is.

On the other hand, if we were provided with an actual
implementation of $f: \{0,1\}^\N   \rightarrow  \{0,1\}^\N$, we
could decide an undecidable problem (namely, $H$) by observing the
trajectories. So there is a discrepancy between the computational
complexity of properties of clopen sets and the actual
possibilities of the machine. This is because we cannot compute
even a single  step of $f$: it is a `non-simulable' system. We
therefore restrict ourselves to systems such that the inverse
image of a clopen set is computable. Note that for instance in
\cite{siegelmann} Siegelmann allows neural networks with
non-recursive weights, leading to a non-computable maps and to
super-Turing capabilities.

\begin{figure}[!h]
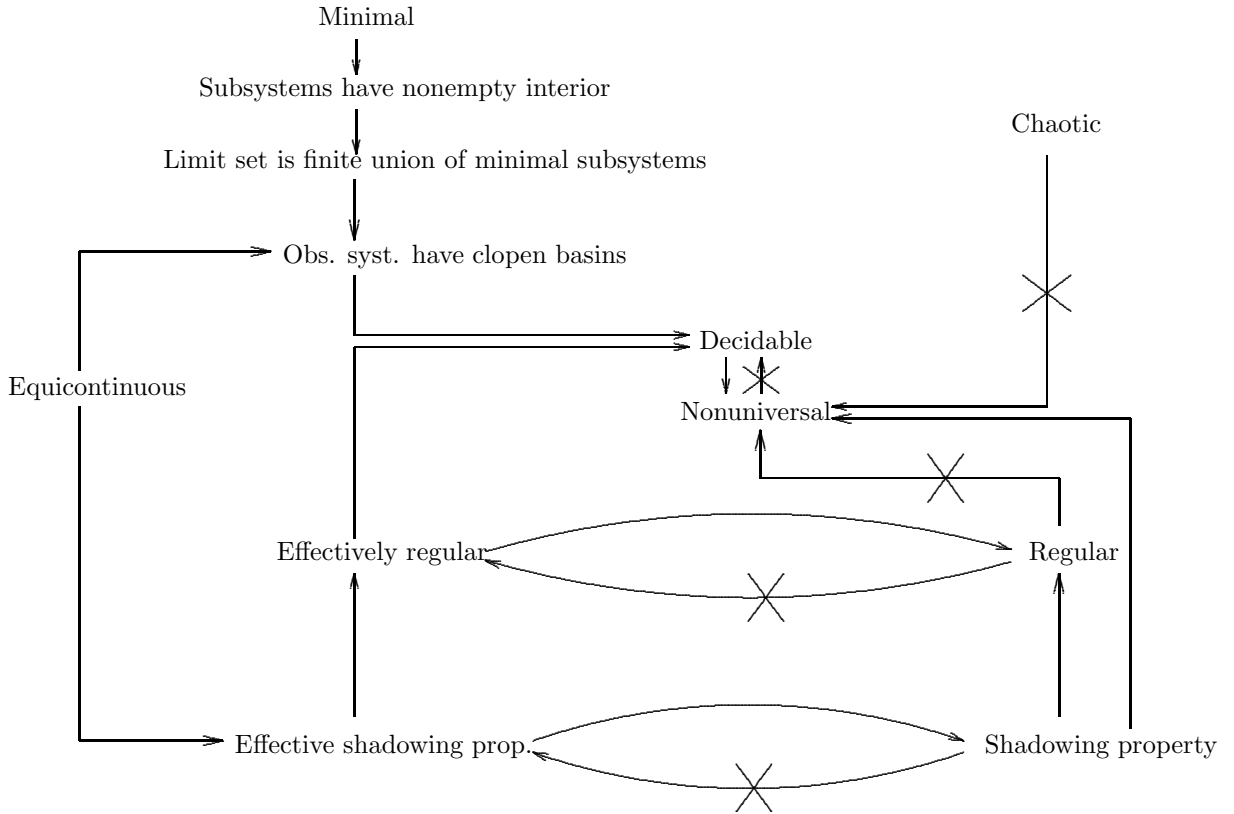

\font\thinlinefont=cmr5
\begingroup\makeatletter\ifx\SetFigFont\undefined%
\gdef\SetFigFont#1#2#3#4#5{%
  \reset@font\fontsize{#1}{#2pt}%
  \fontfamily{#3}\fontseries{#4}\fontshape{#5}%
  \selectfont}%
\fi\endgroup%
\mbox{\beginpicture \setcoordinatesystem units
<1.00000cm,1.00000cm> \unitlength=1.00000cm \linethickness=1pt
\setplotsymbol ({\makebox(0,0)[l]{\tencirc\symbol{'160}}})
\setshadesymbol ({\thinlinefont .}) \setlinear
%
%
\linethickness= 0.500pt \setplotsymbol ({\thinlinefont .}) {\plot
14.542 20.817 15.011 21.169 /
}%
%
%
\linethickness= 0.500pt \setplotsymbol ({\thinlinefont .}) {\plot
14.542 21.169 15.011 20.817 /
}%
%
%
\linethickness= 0.500pt \setplotsymbol ({\thinlinefont .}) {\plot
14.605 17.780 15.073 18.415 /
}%
%
%
\linethickness= 0.500pt \setplotsymbol ({\thinlinefont .}) {\plot
14.605 18.415 15.073 17.780 /
}%
%
%
\put{\SetFigFont{10}{12.0}{\rmdefault}{\mddefault}{\updefault}{Chaotic}%
} [lB] at 18.098 24.289
%
%
\linethickness= 0.500pt \setplotsymbol ({\thinlinefont .}) {\plot
18.256 21.907 18.891 22.384 /
}%
%
%
\linethickness= 0.500pt \setplotsymbol ({\thinlinefont .}) {\plot
18.256 22.384 18.891 21.907 /
}%
%
%
\linethickness= 0.500pt \setplotsymbol ({\thinlinefont .}) {\plot
14.446 15.251 14.912 15.877 /
}%
%
%
\linethickness= 0.500pt \setplotsymbol ({\thinlinefont .}) {\plot
14.446 15.877 14.912 15.251 /
}%
%
%
\linethickness= 0.500pt \setplotsymbol ({\thinlinefont .}) {\plot
16.995 19.367 17.462 20.003 /
}%
%
%
\linethickness= 0.500pt \setplotsymbol ({\thinlinefont .}) {\plot
16.995 20.003 17.462 19.367 /
}%
%
%
\linethickness= 0.500pt \setplotsymbol ({\thinlinefont .})
{%
%
\plot 17.904 18.738 18.098 18.733 17.930 18.828 /
\circulararc 31.906 degrees from 18.098 18.733 center at 14.649
6.502
}%
%
%
\linethickness= 0.500pt \setplotsymbol ({\thinlinefont .})
{%
%
\plot 11.306 18.584 11.113 18.591 11.280 18.494 /
\circulararc 31.645 degrees from 11.113 18.591 center at 14.635
30.906
}%
%
%
\linethickness= 0.500pt \setplotsymbol ({\thinlinefont .})
{%
%
\plot 11.940 16.017 11.748 16.034 11.909 15.929 /
\circulararc 37.961 degrees from 11.748 16.034 center at 14.605
24.342
}%
%
%
\linethickness= 0.500pt \setplotsymbol ({\thinlinefont .})
{%
%
\plot 17.270 16.209 17.462 16.192 17.301 16.298 /
\circulararc 37.961 degrees from 17.462 16.192 center at 14.605
7.885
}%
%
%
\linethickness= 0.500pt \setplotsymbol ({\thinlinefont .})
{\putrule from  9.366 16.510 to  9.366 18.415
%
%
\plot  9.413 18.228  9.366 18.415  9.319 18.228 /
}%
%
%
\linethickness= 0.500pt \setplotsymbol ({\thinlinefont .})
{\putrule from  9.366 18.891 to  9.366 21.431 \putrule from  9.366
21.431 to 13.811 21.431
%
%
\plot 13.624 21.384 13.811 21.431 13.624 21.478 /
}%
%
%
\linethickness= 0.500pt \setplotsymbol ({\thinlinefont .})
{\putrule from 18.574 23.971 to 18.574 20.637 \putrule from 18.574
20.637 to 15.716 20.637
%
%
\plot 15.904 20.684 15.716 20.637 15.904 20.591 /
}%
%
%
\linethickness= 0.500pt \setplotsymbol ({\thinlinefont .})
{%
%
\plot 14.262 21.005 14.309 20.817 14.356 21.005 /
\putrule from 14.309 20.817 to 14.309 21.287
}%
%
%
\linethickness= 0.500pt \setplotsymbol ({\thinlinefont .})
{%
%
\plot 14.823 21.100 14.776 21.287 14.730 21.100 /
\putrule from 14.776 21.287 to 14.776 20.817
}%
%
%
\linethickness= 0.500pt \setplotsymbol ({\thinlinefont .})
{\putrule from 19.685 16.351 to 19.685 20.479 \putrule from 19.685
20.479 to 15.716 20.479
%
%
\plot 15.970 20.542 15.716 20.479 15.970 20.415 /
}%
%
%
\linethickness= 0.500pt \setplotsymbol ({\thinlinefont .})
{%
%
\plot 18.796 18.161 18.733 18.415 18.669 18.161 /
\putrule from 18.733 18.415 to 18.733 16.510
}%
%
%
\linethickness= 0.500pt \setplotsymbol ({\thinlinefont .})
{\putrule from  9.366 23.654 to  9.366 22.860
%
%
\plot  9.303 23.114  9.366 22.860  9.430 23.114 /
}%
%
%
\linethickness= 0.500pt \setplotsymbol ({\thinlinefont .})
{\putrule from  9.366 22.384 to  9.366 21.590 \putrule from  9.366
21.590 to 13.811 21.590
%
%
\plot 13.624 21.543 13.811 21.590 13.624 21.637 /
}%
%
%
\linethickness= 0.500pt \setplotsymbol ({\thinlinefont .})
{\putrule from  5.715 21.114 to  5.715 22.701 \putrule from  5.715
22.701 to  8.255 22.701
%
%
\plot  8.001 22.638  8.255 22.701  8.001 22.765 /
}%
%
%
\linethickness= 0.500pt \setplotsymbol ({\thinlinefont .})
{\putrule from  5.715 20.637 to  5.715 16.192 \putrule from  5.715
16.192 to  7.620 16.192
%
%
\plot  7.366 16.129  7.620 16.192  7.366 16.256 /
}%
%
%
\linethickness= 0.500pt \setplotsymbol ({\thinlinefont .})
{\putrule from  9.392 25.525 to  9.392 25.057
%
%
\plot  9.345 25.244  9.392 25.057  9.438 25.244 /
}%
%
%
\linethickness= 0.500pt \setplotsymbol ({\thinlinefont .})
{\putrule from  9.392 24.589 to  9.392 24.001
%
%
\plot  9.345 24.188  9.392 24.001  9.438 24.188 /
}%
%
%
\linethickness= 0.500pt \setplotsymbol ({\thinlinefont .})
{\putrule from 18.733 19.050 to 18.733 19.685 \putrule from 18.733
19.685 to 14.764 19.685 \putrule from 14.764 19.685 to 14.764
20.320
%
%
\plot 14.827 20.066 14.764 20.320 14.700 20.066 /
}%
%
%
\put{\SetFigFont{10}{12.0}{\rmdefault}{\mddefault}{\updefault}{Decidable}%
} [lB] at 13.957 21.404
%
%
\put{\SetFigFont{10}{12.0}{\rmdefault}{\mddefault}{\updefault}{Nonuniversal}%
} [lB] at 13.702 20.464
%
%
\put{\SetFigFont{10}{12.0}{\rmdefault}{\mddefault}{\updefault}{Regular}%
} [lB] at 18.332 18.591
%
%
\put{\SetFigFont{10}{12.0}{\rmdefault}{\mddefault}{\updefault}{Shadowing property}%
} [lB] at 17.748 16.034
%
%
\put{\SetFigFont{10}{12.0}{\rmdefault}{\mddefault}{\updefault}{Effective shadowing prop.}%
} [lB] at  7.779 16.034
%
%
\put{\SetFigFont{10}{12.0}{\rmdefault}{\mddefault}{\updefault}{Effectively regular}%
} [lB] at  8.335 18.591
%
%
\put{\SetFigFont{10}{12.0}{\rmdefault}{\mddefault}{\updefault}{Obs. syst. have clopen basins}%
} [lB] at  8.414 22.543
%
%
\put{\SetFigFont{10}{12.0}{\rmdefault}{\mddefault}{\updefault}{Equicontinuous}%
} [lB] at  4.763 20.796
%
%
\put{\SetFigFont{10}{12.0}{\rmdefault}{\mddefault}{\updefault}{Subsystems have nonempty interior}%
} [lB] at  7.303 24.765
%
%
\put{\SetFigFont{10}{12.0}{\rmdefault}{\mddefault}{\updefault}{Minimal}%
} [lB] at  8.890 25.718
%
%
\put{\SetFigFont{10}{12.0}{\rmdefault}{\mddefault}{\updefault}{Limit set is finite union of minimal subsystems}%
} [lB] at  6.826 23.812 \linethickness=0pt \putrectangle corners
at  4.763 25.940 and 20.449 15.225
\endpicture}
  \caption{Summary of the results. Arrows read `implies',
  crossed arrows read `does not imply'.} \label{figconclusion}
\end{figure}

\section{Conclusions and future work}

\label{future} \label{sectconcl}

We provided a definition of decidability and universality for a
symbolic systems, and established some links between decidability
and the dynamical properties of the system. We also constructed a
chaotic system that is universal. These results are summed up in
Figure \ref{figconclusion}. We have already formulated some open
problems. Is there a cellular automaton that is chaotic and
universal? Do undecidable system have infinitely many disjoint
subsystems? And many more questions are yet to be solved. For
instance, can we find sufficient conditions of universality? Which
simplicity criteria proposed in \cite{kurka_simplicity} are
sufficient conditions for decidability? Are the Game of Life and
the automaton 110 universal for our definition? Can a linear
cellular automaton be universal?

It also remains to extend  the definitions and results to systems
in $\R^n$ in discrete time or even continuous time. The resulting
definition of universality could then be compared to existing
definitions, for instance
\cite{siegelmann__fishman_analog_dissipative,
bournez_cosnard,orponen_survey,moore_recogn}. Then, results such
as those of Section \ref{sectnecesscond} could hopefully be
adapted. For instance, are minimal systems capable of universal
computation? Such results could then be applied to physical
systems. What systems that can be found in Nature are able to
compute? For instance, hyperbolic dynamical systems are known to
have the effective shadowing property. This would suggest that
hyperbolic systems are not universal.

A theory a computational complexity could also be investigated.
What problems  can be solved in polynomial time with a
discrete-time dynamical system? Can we formulate a `P$\neq$NP'
conjecture? See \cite{BenHurSiegelmannFishman02,
siegelmann__fishman_analog_dissipative} for theories of complexity
in analog computation.

\section{Acknowledgements}
We warmly thank Eugene Asarin (who suggested Proposition
\ref{prop_analytic_hier}), Anahi Gajardo, Olivier Bournez, Pascal
Koiran and Cris Moore for fruitful discussions and comments. This
paper presents research results of the Belgian Programme on
Interuniversity Attraction Poles, initiated by the Belgian Federal
Science Policy Office. The scientific responsibility rests with
its authors. J.-Ch. D. holds a FNRS fellowship (Belgian Fund for
Scientific Research).


\begin{thebibliography}{10}

\bibitem{asarin_bouajjani}
E.~Asarin and A.~Bouajjani.
\newblock Perturbed turing machines and hybrid systems.
\newblock In {\em Proceedings of the 16th Annual {IEEE} Symposium on Logic in
  Computer Science ({LICS}-01)}, pages 269--278, Los Alamitos, CA, June ~16--19
  2001. IEEE Computer Society.

\bibitem{asarin_pnu_maler_95}
E.~Asarin, O.~Maler, and A.~Pnueli.
\newblock Reachability analysis of dynamical systems having piecewise-constant
  derivatives.
\newblock {\em Theoretical Computer Science}, 138(1):35--65, 1995.
\newblock Special issue on hybrid systems.

\bibitem{banks_sensitiv_in_chaos}
J.~Banks, J.~Brooks, G.~Cairns, G.~Davis, and P.~Stacey.
\newblock On {D}evaney's definition of chaos.
\newblock {\em American Mathematics Monthly}, 99:332--334, 1992.

\bibitem{BenHurSiegelmannFishman02}
A.~Ben-Hur, H.~T. Siegelmann, and S.~Fishman.
\newblock A theory of complexity for continuous time systems.
\newblock {\em Journal of Complexity}, 18(1):51--86, 2002.

\bibitem{bournez_cosnard}
O.~Bournez and M.~Cosnard.
\newblock On the computational power of dynamical systems and hybrid systems.
\newblock {\em Theoretical Computer Science}, 168:417--459, 1996.

\bibitem{conway_collatz}
J.~H. Conway.
\newblock Unpredictable iterations.
\newblock In {\em Proceedings of the Number Theory Conference (Univ. Colorado,
  Boulder, Colo., 1972)}, pages 49--52, Boulder, Colo., 1972. Univ. Colorado.

\bibitem{crutchfield_onsetchaos}
J.~P. Crutchfield and K.~Young.
\newblock Computation at the onset of chaos.
\newblock In W.~Zurek, editor, {\em Complexity, Entropy and the Physics of
  Information}, pages 223--269. Addison-Wesley, 1989.

\bibitem{davis_universality}
M.~D. Davis.
\newblock A note on universal {T}uring machines.
\newblock In C.E. Shannon and J.~McCarthy, editors, {\em Automata Studies},
  pages 167--175. Princeton University Press, 1956.

\bibitem{delvenne_kurka_blondel}
J.-Ch. Delvenne, P.~K\r{u}rka, and V.~D. Blondel.
\newblock Computational universality in symbolic dynamical systems.
\newblock In M.~Margenstern, editor, {\em MCU 2004}, number 3354 in Lecture
  Notes in Computer Science, pages 104--115. Springer-Verlag, 2005.

\bibitem{devaney}
R.~Devaney.
\newblock {\em An Introduction to Chaotic Dynamical Systems}.
\newblock Addison-Wesley, 1989.

\bibitem{durand_roka_gameoflife}
B.~Durand and Z.~R\'{o}ka.
\newblock The {G}ame of {L}ife: universality revisited.
\newblock In M.~Delorme and J.~Mazoyer, editors, {\em Cellular Automata: a
  Parallel Model}, volume 460 of {\em Mathematics and its Applications}, pages
  51--74. Kluwer Academic Publishers, 1999.

\bibitem{Fredk_Toffo_billiard}
Edward Fredkin and Tommaso Toffoli.
\newblock Conservative logic.
\newblock {\em International Journal of Theoretical Physics}, 21(3-4):219--253,
  1981/82.
\newblock Physics of computation, Part I (Dedham, Mass., 1981).

\bibitem{gacs97reliable}
Peter G{\'a}cs.
\newblock Reliable cellular automata with self-organization.
\newblock In {\em 38th Annual Symposium on Foundations of Computer Science},
  pages 90--99, Miami Beach, Florida, 20--22 October 1997. IEEE.

\bibitem{hemmerling}
A.~Hemmerling.
\newblock Effective metric spaces and representations of the reals.
\newblock {\em Theoretical Computer Science}, 284(2):347--372, 2002.

\bibitem{modelchecking1999}
E.~M.~Clarke Jr., O.~Grumberg, and D.~A. Peled.
\newblock {\em Model checking}.
\newblock MIT Press, 1999.

\bibitem{rogers_book}
H.~Rogers Jr.
\newblock {\em Theory of recursive functions and effective computability}.
\newblock MIT Press, Cambridge, MA, second edition, 1987.

\bibitem{kaivola_thesis}
R.~Kaivola.
\newblock {\em Using automata to characterise fixed point temporal logics}.
\newblock PhD thesis, University of Edinburgh, 1996.

\bibitem{koiran_recur_networks_96}
P.~Koiran.
\newblock A family of universal recurrent networks.
\newblock {\em Theoretical Computer Science}, 168(2):473--480, 1996.
\newblock Universal machines and computations (Paris, 1995).

\bibitem{koiran_piecewise_lin_maps_94}
P.~Koiran, M.~Cosnard, and M.~Garzon.
\newblock Computability with low-dimensional dynamical systems.
\newblock {\em Theoretical Computer Science}, 132(1-2):113--128, 1994.

\bibitem{koiran_moore_99}
P.~Koiran and Cr. Moore.
\newblock Closed-form analytic maps in one and two dimensions can simulate
  universal {T}uring machines.
\newblock {\em Theoretical Computer Science}, 210(1):217--223, 1999.

\bibitem{kurka_simplicity}
P.~K\r{u}rka.
\newblock Simplicity criteria for dynamical systems.
\newblock {\em Analysis of Dynamical and Cognitive Systems}, pages 189--225,
  1993.

\bibitem{kurka_equicont_cellaut}
P.~K\r{u}rka.
\newblock Languages, equicontinuity and attractors in cellular automata.
\newblock {\em Ergodic Theory \& Dynamical Systems}, 17:417--433, 1997.

\bibitem{kurka_topo_TM}
P.~K\r{u}rka.
\newblock On topological dynamics of {T}uring machines.
\newblock {\em Theoretical Computer Science}, 174(1-2):203--216, 1997.

\bibitem{kurka_universality}
P.~K\r{u}rka.
\newblock Zero-dimensional dynamical systems, formal languages, and
  universality.
\newblock {\em Theory of Computing Systems}, 32(4):423--433, 1999.

\bibitem{kurka_book}
P.~K\r{u}rka.
\newblock {\em Topological and Symbolic Dynamics}, volume~11 of {\em Cours
  Spécialisés}.
\newblock Société Mathématique de France, 2004.

\bibitem{langton_edgeofchaos}
C.~G. Langton.
\newblock Computation at the edge of chaos.
\newblock {\em Physica D}, 42:12--37, 1990.

\bibitem{maass_orponen}
W.~Maass and P.~Orponen.
\newblock On the effect of analog noise in discrete-time analog computations.
\newblock {\em Neural Computation}, 10(5):1071--1095, 1998.

\bibitem{minsky1961}
M.~L. Minsky.
\newblock Recursive unsolvability of {P}ost's problem of ``tag'' and other
  topics in theory of {T}uring machines.
\newblock {\em Annals of Mathematics (2)}, 74:437--455, 1961.

\bibitem{edgeofchaos_reexamination}
M.~Mitchell, P.~T. Hraber, and J.~P. Crutchfield.
\newblock Dynamic computation, and the ``edge of chaos'': {A} re-examination.
\newblock In G.~Cowan, D.~Pines, and D.~Melzner, editors, {\em Complexity:
  Metaphors, Models, and Reality}, Santa Fe Institute Proceedings, Volume 19,
  pages 497--513. Addison-Wesley, 1994.
\newblock Santa Fe Institute Working Paper 93-06-040.

\bibitem{moore_unpred_and_undec}
Cr. Moore.
\newblock Unpredictability and undecidability in dynamical systems.
\newblock {\em Physical Review Letters}, 64(20):2354--2357, 1990.

\bibitem{moore_generalizedshifts}
Cr. Moore.
\newblock Generalized shifts: Unpredictability and undecidability in dynamical
  systems.
\newblock {\em Nonlinearity}, 4:199--230, 1991.

\bibitem{moore_recogn}
Cr. Moore.
\newblock Dynamical recognizers: real-time language recognition by analog
  computers.
\newblock {\em Theoretical Computer Science}, 201:99--136, 1998.

\bibitem{moore_auckland}
Cr. Moore.
\newblock Finite-dimensional analog computers: Flows, maps, and recurrent
  neural networks.
\newblock In C.S. Calude, J.~Casti, and M.~J. Dinneen, editors, {\em
  Unconventional Models of Computation}. Springer-Verlag, 1998.

\bibitem{orponen_survey}
P.~Orponen.
\newblock A survey of continuous-time computation theory.
\newblock In D.-Z. Du and K.-I Ko, editors, {\em Advances in Algorithms,
  Languages, and Complexity}, pages 209--224. Kluwer Academic Publishers, 1997.

\bibitem{perrinpin}
D.~Perrin and J.-E. Pin.
\newblock {\em Infinite Words (Automata, Semigroups, Logic and Games)}, volume
  141 of {\em Pure and Applied Mathematics}.
\newblock Elsevier, 2004.

\bibitem{siegelmann}
H.~T. Siegelmann.
\newblock {\em Neural Networks and Analog Computation: Beyond the {T}uring
  Limit}.
\newblock Progress in Theoretical Computer Science. Springer-Verlag, 1999.

\bibitem{siegelmann__fishman_analog_dissipative}
H.~T. Siegelmann and S.~Fishman.
\newblock Analog computation with dynamical systems.
\newblock {\em Physica D}, 120:214--235, 1998.

\bibitem{sutner_intermediateCA}
K.~Sutner.
\newblock Cellular automata and intermediate degrees.
\newblock {\em Theoretical Computer Science}, 296(2):365--375, 2003.

\bibitem{turing}
A.~M. Turing.
\newblock On computable numbers, with an application to the
  {E}ntscheidungsproblem.
\newblock {\em Proceedings of the London Mathematical Society}, 42(2):230--265,
  1936.

\bibitem{weihrauch}
K.~Weihrauch.
\newblock {\em Computable Analysis}.
\newblock Springer-Verlag, 2000.

\bibitem{wolfram}
S.~Wolfram.
\newblock {\em A new kind of science}.
\newblock Wolfram Media, Inc., Champaign, IL, 2002.

\end{thebibliography}
\end{document}